\documentclass[reprint,twocolumn,superscriptaddress,preprintnumbers,amsmath,amssymb,aps,prd,floatfix]{revtex4-2}

\usepackage[utf8]{inputenc}

\usepackage{mathtools}
\usepackage{amsfonts}
\usepackage{mathrsfs}
\usepackage{bm}
\usepackage[normalem]{ulem}
\usepackage{bbold}
\usepackage{braket}
\usepackage{slashed}
\usepackage{dsfont}
\usepackage{float}
\usepackage{dcolumn}
\usepackage{amssymb,amsmath}
\usepackage{titlesec}

\usepackage{graphicx}
\usepackage{color}
\usepackage{array}

\usepackage{placeins}
\usepackage{booktabs}
\usepackage{orcidlink}

\usepackage{xspace}
\usepackage{hyperref}
\usepackage[nameinlink]{cleveref}
\usepackage{bookmark}
\usepackage{units}

\setkeys{Gin}{width=0.48\textwidth}

\newcommand{\tr}{{\text{tr}}}

\usepackage{booktabs}
\usepackage{multirow}

\newcolumntype{C}{>{$}c<{$}}
\AtBeginDocument{
	\heavyrulewidth=.08em
	\lightrulewidth=.05em
	\cmidrulewidth=.03em
	\belowrulesep=.65ex
	\belowbottomsep=0pt
	\aboverulesep=.4ex
	\abovetopsep=0pt
	\cmidrulesep=\doublerulesep
	\cmidrulekern=.5em
	\defaultaddspace=.5em
}

\newcommand{\LEGO}{LEGO\textsuperscript{\textregistered}}

\graphicspath{{./figures/}}

\usepackage{xifthen}
\usepackage{xcolor}

\hypersetup{
	colorlinks,
	linkcolor={blue!75!black},
	citecolor={blue!75!black},
	urlcolor={blue!75!black}, 
	pdftitle={},
	pdfauthor={},
	pdfkeywords={}
	{} 
	bookmarksopen=true,
	bookmarksopenlevel=2,
	bookmarksnumbered=true
}

\begin{document}

\title{Strangeness neutrality and the QCD phase diagram}

\author{Wei-jie Fu \,\orcidlink{0000-0002-5647-5246}}
\affiliation{School of Physics, Dalian University of Technology, Dalian, 116024,
		P.R. China}

\author{Chuang Huang \, \orcidlink{0000-0001-7873-8978}}
\affiliation{Institut f\"ur Theoretische Physik, Universit\"at Heidelberg, Philosophenweg 16, 69120 Heidelberg, Germany}
  
\author{Jan M. Pawlowski \,\orcidlink{0000-0003-0003-7180}}
\affiliation{Institut f\"ur Theoretische Physik, Universit\"at Heidelberg, Philosophenweg 16, 69120 Heidelberg, Germany}
	\affiliation{ExtreMe Matter Institute EMMI, GSI, Planckstra{\ss}e 1, D-64291 Darmstadt, Germany}

\author{Fabian Rennecke \,\orcidlink{0000-0003-1448-677X}}
\affiliation{Institut f\"ur Theoretische Physik, Justus-Liebig-Universit\"at Gie\ss en, 35392 Gie\ss en, Germany}
\affiliation{Helmholtz Research Academy Hesse for FAIR, Campus Gie\ss en, 35392 Gie\ss en, Germany}

\author{Rui Wen \,\orcidlink{0000-0002-1319-1331}}
\email{rwen@ucas.ac.cn}
\affiliation{School  of  Nuclear  Science  and  Technology, University  of  Chinese  Academy  of  Sciences,  Beijing,  P.R.China  100049}
	
\author{Shi Yin \,\orcidlink{0000-0001-5279-6926}}
\affiliation{Institut f\"ur Theoretische Physik, Justus-Liebig-Universit\"at Gie\ss en, 35392 Gie\ss en, Germany}

\begin{abstract}

We map out the phase structure of $N_f=2+1$ flavour QCD at strangeness neutrality with functional QCD. We find a critical end point at $(T_{\rm CEP},\mu_{B,{\rm CEP}})|_{n_S=0} = (92, 696)$\,MeV. The computation is done with the functional renormalisation group, and we systematically improve on previous works, hence reducing the systematic error significantly. Our results pass relevant QCD benchmarks: they agree well with and corroborate the QCD phase structure from functional QCD results at vanishing strangeness chemical potential. Moreover, they agree  well with lattice QCD results at vanishing chemical potential. Specifically, the ratio of the second order curvature coefficient $\kappa_2$ agrees with that obtained from lattice computations, $\kappa_2(n_S=0)/\kappa_2(\mu_S=0)=0.897(20)$.   

\end{abstract}

\maketitle

\emph{Introduction.--} 
The phase structure of QCD has attracted a lot of attention in recent years. Particular attention has been devoted to the regime with the onset of new phases (ONP), either a critical end point that hallmarks a first order regime or regimes with instabilities or moats \cite{Fu:2019hdw, Pisarski:2021qof, Pawlowski:2025jpg}. Both experiment and theory have made great efforts and significant progress in the last decades. The Beam Energy Scan (BES) program at the Relativistic Heavy Ion Collider (RHIC) covers the region of collision energies from 200\,GeV to 3.0\,GeV, including the onset regime of new phases, see \cite{Fu:2019hdw, Gao:2020fbl, Gunkel:2021oya, Pawlowski:2025jpg}. 

Cumulants of net-proton distributions, e.g.~\cite{STAR:2020tga, STAR:2021iop, STAR:2021rls, STAR:2021fge, STAR:2022vlo, STAR:2025zdq}, fluctuations of strangeness and electric charge, e.g.~\cite{Adamczyk:2017wsl, Pandav:2020uzx}, correlations of conserved charges \cite{STAR:2019ans}, the light nuclei production \cite{STAR:2022hbp, STAR:2023uxk}, and the spin alignment \cite{STAR:2022fan}, could provide signals for the existence of the CEP, see also \cite{Luo:2017faz, Pandav:2022xxx, Aparin:2022jok, Zhang:2026dny} for  reviews of  recent results by the STAR collaboration. Furthermore, several other heavy-ion experiments are planned and will be devoted to further explore the QCD phase diagram at high baryon chemical potential, such as the CBM \cite{CBM:2016kpk} and CEE+ \cite{Ruan:2018fpo} experiments.

The theoretical description of the QCD phase structure has also advanced significantly. Recently, the large density regime with $\sqrt{s}\lesssim 10$\,GeV has been mapped out with functional 2+1 flavour QCD, \cite{Fu:2019hdw, Gao:2020fbl, Gunkel:2021oya, Pawlowski:2025jpg}. These studies either used identical quark chemical potentials, $\mu_q=\mu_B/3$ ($\mu_S=0$), or only light quark chemical potentials $\mu_l=\mu_B/3$ and $\mu_s=0$. Importantly, these computations have narrowed down the location of the CEP in these cases, or more generally, that of the onset of new phases: $(T_{\rm CEP},\mu_{B,{\rm CEP}}) \approx (110\pm 10,630\pm 30)$\,MeV. This corresponds to collision energies of about 3-4\,GeV per nucleon, putting it within reach of fixed-target heavy-ion collision experiments. 
In addition, thermodynamic observables and fluctuations of conserved charges are available from functional methods \cite{Fu:2021oaw, Fu:2023lcm, Lu:2023mkn, Lu:2025cls, Chen:2025vwl, Lu:2026ezr} at finite density, providing guidance for future heavy-ion experiments;
see \cite{Fischer:2026uni} for an overview of recent developments. 
Owing to the sign problem, finite $\mu_B$ is not directly accessible by lattice QCD. Still, estimates of the CEP location based on extrapolations of lattice data agree with the predictions from functional methods. Recent estimates are obtained from the reconstruction of Yang-Lee edge singularities \cite{Basar:2023nkp, Clarke:2024ugt, Adam:2025phc} and contours of constant entropy density \cite{Shah:2024img, Borsanyi:2025dyp}.

In this work, we provide the first functional QCD analysis of the QCD phase structure at strangeness neutrality, $n_S=0$.  Specifically, we accommodate the dynamics of the scalar and pseudoscalar nonets, hence including open strange mesons such as the kaons and the flavour-octet $\eta$-meson. This allows us to map out the chiral crossover line, which ends in a chiral CEP at baryon chemical potential $\mu_B\approx 700$\,MeV. \\[-1.5ex] 

%
\begin{figure*}[t]
	\includegraphics[width=0.44\textwidth]{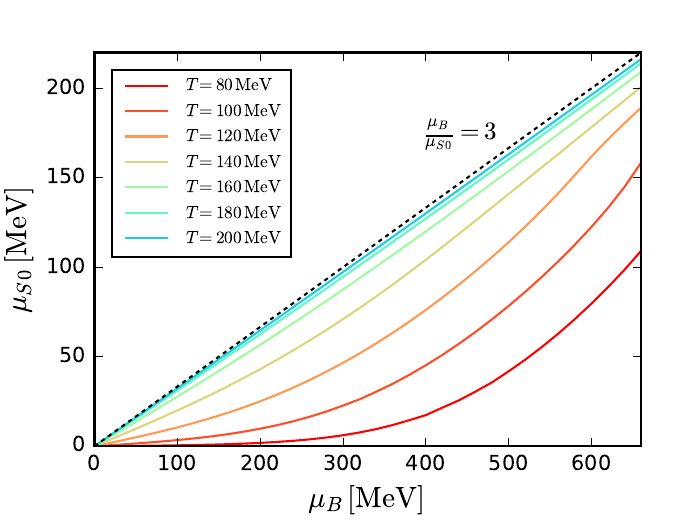}\hspace{.5cm}
	\includegraphics[width=0.52\textwidth]{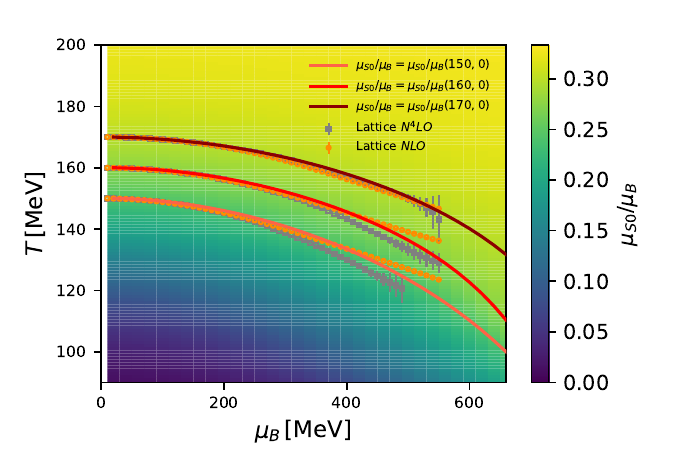}
	\caption{Left: Strangeness chemical potential $\mu_{S0}$ at strangeness neutrality, \labelcref{eq:muS0}, as a function of the baryon chemical potential $\mu_{B}$. The free quark gas limit $\mu_B/\mu_{S}=3$ is also plotted for comparison (black dashed line).\\
		Right: Heat map of the ratio $\mu_{S0}/\mu_B$ in the $T - \mu_B$ plane. We also show the the contours of constant $\mu_{S0}/\mu_{B}(T,\mu_B)$ for $T(\mu_B=0)=(150, 160, 170)$\,MeV in comparison with the lattice QCD results \cite{Borsanyi:2025kiv}.
	}\label{fig:muS0}
\end{figure*}
%

\emph{Functional QCD at strangeness neutrality.--}
The computations and results in the present work are achieved within the fRG-approach to QCD at finite temperature and density, set up in \cite{Fu:2019hdw} and extended in \cite{Fu:2024rto, Ihssen:2024miv, Pawlowski:2025jpg}. In these works the effective action of QCD is obtained by solving its flow equation from a large infrared cutoff scale $k=\Lambda_\textrm{UV}$ with a perturbative effective action $\Gamma_\Lambda$ 
down to vanishing cutoff, $k=0$ with the full quantum effective action $\Gamma=\Gamma_{k=0}$. The discussion of our expansion scheme and details on the flow equations for correlation functions and their solutions is deferred to the Supplement. 

The effective action of QCD, evaluated on the solution of the equations of motion, is simply the grand potential $\Omega(T,\mu_B,\mu_S)$. Here, $\mu_B$ is the baryon chemical potential and the strangeness chemical potential $\mu_S$ are given by  
\begin{align} 
	\mu_l=\frac{\mu_B}{3}\,,\qquad \mu_S=\mu_l-\mu_s\,, 
\label{eq:muB-muS}
\end{align} 
with the quark chemical potentials $\mu_{q_i}$ with $q=(u,d,s)$. 
In \labelcref{eq:muB-muS} we have identified the $u,d$ chemical potentials $\mu_{u}=\mu_d=\mu_l$, which implies a vanishing electric charge chemical potential. For more details see the Supplement.

The grand potential $\Omega$ is directly related to the effective action, evaluated on the equations of motion,
\begin{align}
	\Omega(T,\mu_B,\mu_S)=\frac{T}{V}\Bigl[ \Gamma(T,\mu_B,\mu_S)-\Gamma(0,0,0)\Bigr]\,,
	\label{eq:GrandPotential}
\end{align}
with the temperature $T$ and spatial volume $V$. The generalised susceptibilities are defined as 
\begin{align}
	\chi_{ij}^{BS}=&\frac{\partial^{i+j} (p/T^4)}{\partial(\mu_{B}/T)^{i}\partial(\mu_{S}/T)^{j}}\,,  
\label{eq:generalised_susceptibilities}
\end{align}
with the pressure $p=-\Omega$ in the absence of volume fluctuations. 
In particular, the strangeness density reads 
\begin{align} 
n_S=\chi_{1}^{S} \,.
\label{eq:nS} 
\end{align} 
The initial state of a heavy ion collision (HIC) is strangeness neutral, $n_S=0$. This property persists throughout the timeline of a HIC as strangeness is a conserved charge in QCD. For any given $T$ and $\mu_B$, this implies a certain strangeness chemical potential, 
\begin{align} 
	\mu_{S0}= \mu_{S}(T,\mu_B)\,,\quad \textrm{with} \quad n_S(T,\mu_B,\mu_{S0}) =0 \,.
\label{eq:muS0}
\end{align}
The resulting $T$--$\mu_B$ phase diagram is a hypersurface in the space spanned by $\{T,\mu_B,\mu_S\}$. In \Cref{fig:muS0}, we show our results for $\mu_{S0}$ as a function of $\mu_B$ for various temperatures (left), and a heat map of the ratio $\mu_{S0}/\mu_{B}$ in the $T - \mu_B $ plane (right). 

%
\begin{figure*}[t]
	\includegraphics[width=0.98\textwidth]{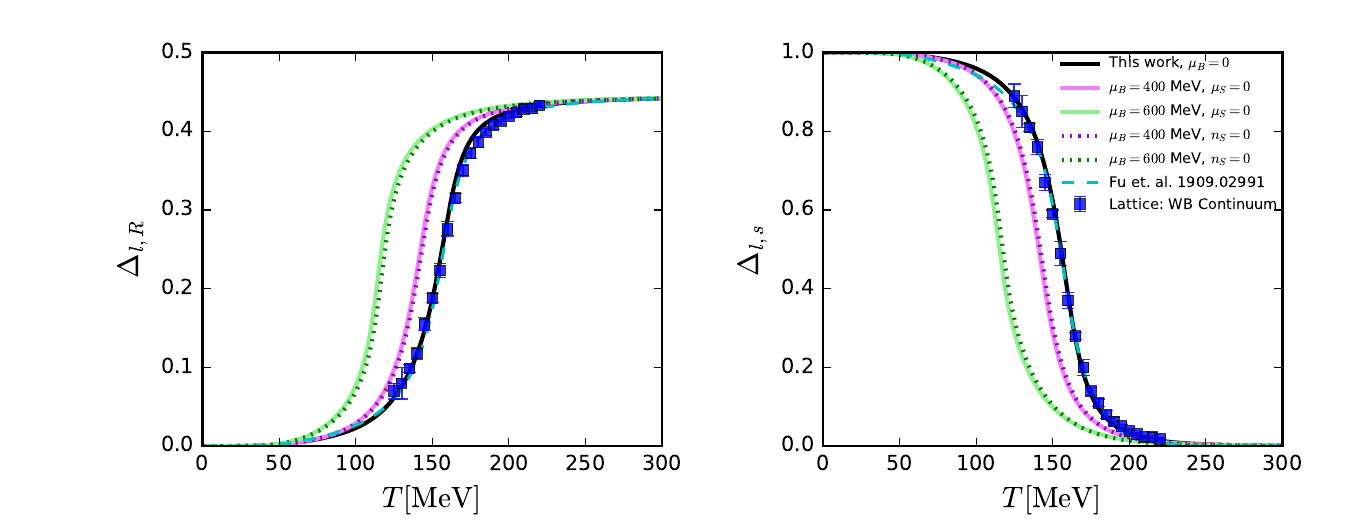}
	\caption{Renormalised light quark chiral condensate $\Delta_{l,R}$ (left panel) and the reduced condensate $\Delta_{l,s}$ (right panel) as functions of temperature. We also show functional QCD results from \cite{Fu:2019hdw}, and lattice QCD results from the Wuppertal-Budapest collaboration \cite{Borsanyi:2010bp} at vanishing chemical potential. }\label{fig:Delta}
\end{figure*}
%
These results for $\mu_{S0}$ are obtained within a numerical resolution of $n_S=0$ in \labelcref{eq:muS0}. Alternatively one may use
\begin{align}
	\frac{\partial \mu_{S}}{\partial \mu_B}=-\frac{\chi^{BS}_{11}(T,\mu_B,\mu_{S})}{\chi^S_2(T,\mu_B,\mu_{S})}\,, 
\label{eq:muS0Flucs}
\end{align}
see e.g.~\cite{Bellwied:2015rza, Fu:2018swz, Wen:2019ruz}. 
The second-order susceptibilities $\chi^{BS}_{11}$ and $\chi^{S}_{2}$ are defined in \labelcref{eq:generalised_susceptibilities}. Results for \labelcref{eq:muS0Flucs} as well as a comparison of general susceptibilities with the respective lattice results will be done in \cite{Wen2026B-S}. Here we only note that \labelcref{eq:muS0Flucs} provides us with a direct relation between strangeness neutrality and baryon-strangeness correlations, $\chi^{BS}_{11}/\chi^{S}_{2}$. The latter are sensitive probes of the nature of QCD matter \cite{Koch:2005vg}. Hence, \labelcref{eq:muS0Flucs} carries the relation between quark number conservation and the QCD phase structure \cite{Fu:2018swz}. 

This allows us to discuss some important properties of the curves $\mu_{S0}(\mu_B)$ shown in the left panel in \Cref{fig:muS0}: \\[-2ex]
 
For high temperatures and baryon chemical potential, the system approaches the free quark gas limit. There, strangeness is dominated by strange quarks and hence we find $\mu_s=\mu_B/3 - \mu_{S}=0$.

For low temperatures and baryon chemical potential, strangeness is dominated by open strange mesons, making the baryon-strangeness correlation in \labelcref{eq:muS0Flucs} very small. There,  $\mu_{S}$ is much smaller than $\mu_B/3$. 

As the baryon chemical potential rises, strange baryons become increasingly relevant, leading to increasing $\chi^{BS}_{11}$ and, consequently, larger $\mu_{S}$. In the deconfined phase, the free quark limit is approached. \\[-2ex]

We close this assessment of strangeness neutrality with a discussion of the lines of constant ratios $\mu_{S0}/\mu_{B}$. They are shown in the right panel of \Cref{fig:muS0}, starting at $T(\mu_B=0)=150,160,170$\,MeV. Our results are in quantitative agreement with the lattice QCD results of \cite{Borsanyi:2025kiv} for small baryon chemical potential for $\mu_B/T \lesssim 4/3$. Note that this estimate has to be taken with a grain of salt as for different $T(\mu_B=0)$ we expect different convergence ranges for the Taylor expansion underlying the lattice results. In any case, the lattice shows an increasingly worse convergence for $\mu_B/T \gtrsim 4/3$ as well as an increasing deviation from our results. At present, the results achieved here are the only (functional) QCD results at strangeness neutrality for large baryon chemical potential.   \\[-1.5ex]

\emph{Chiral condensates and thermal susceptibilities.--} 
We proceed with the preparation of the computation of the chiral phase structure of QCD at strangeness neutrality. The chiral crossover temperature is computed from the thermal susceptibilities of variants of the chiral condensate. The chiral condensate of a given quark flavour $q=(l,s)$ and $l=(u,d)$ is defined as
\begin{align}
\Delta_{q_i}=m_{q_i}^0\frac{T}{V}\int_x\langle {\bar q}_i (x)q_i(x)\rangle\,,
\label{eq:ChiralCondensate}
\end{align}
where no sum over the flavour index $i$ is taken. If computed from its fRG-flow, \labelcref{eq:ChiralCondensate} is finite as the flow is finite. Put differently, the renormalisation is built-in by construction. Moreover, in the fRG-approach with emergent composites, the chiral condensates are proportional to the solutions $\sigma^\textrm{EoM}, \sigma_s^\textrm{EoM}$ of the equation of motion for the scalar modes: $\Delta_{l} \propto \sigma^\textrm{EoM}$ with the 
light scalar mode $\sigma^\textrm{EoM} \propto \langle \bar l l\rangle_r $, and  $\Delta_{s} \propto \sigma^\textrm{EoM}_s$ with the strange scalar mode $\sigma^\textrm{EoM}_s \propto \langle \bar s s\rangle_r $. Here the subscript ${}_r$ indicates that these are renormalised expectation values. For a comprehensive discussion see \cite{Fu:2019hdw}. 

On the lattice and in other functional approaches, \labelcref{eq:ChiralCondensate} has to be renormalised. This is commonly done or rather circumvented by computing either the \textit{renormalised} or the \textit{reduced} condensates. For the sake of a  direct comparison we shall also compute these condensates. The renormalised condensates are defined as 
\begin{align}
\Delta_{q_i,R}=\frac{1}{\mathcal{N}_R}[\Delta_{q_i}(T,\mu_q)-\Delta_{q_i}(0,0)]\,,
\label{eq:DeltaR}
\end{align}
where $\mathcal{N}_R$ is a normalisation constant at our disposal. The reduced condensate is defined by  
\begin{align}
\Delta_{l,s}&\equiv\frac{\Delta_{l}(T,\mu_B)-\Big( \frac{m_l^0}{m_s^0} \Big)^2 \Delta_s(T,\mu_B)}{\Delta_{l}(0,0)-\Big( \frac{m_l^0}{m_s^0} \Big)^2 \Delta_s(0,0)}\,, 
\label{eq:Deltals}
\end{align}
which eliminates the normalisation ${\cal N}_R$ and the volume factor in  \labelcref{eq:ChiralCondensate}. 

In \Cref{fig:Delta}, we show both the renormalised light chiral condensate $\Delta_{l,R}$ and the reduced condensate $\Delta_{l,s}$ as functions of temperature at vanishing and finite baryon chemical potential and strangeness neutrality, $n_S=0$.  For vanishing chemical potential,  $\mu_B=0=\mu_S$, we compare our results with lattice calculations \cite{Borsanyi:2010bp}. We also show the results for $\mu_S=0$ from the previous functional QCD work \cite{Fu:2019hdw}. 
The comparison documents the quantitative agreement of the present results for the chiral condensates with lattice and the previous functional QCD results at $\mu_B=0$. 

In the following we consider the chiral crossover temperature $T_c$ given by the peak position of the thermal susceptibility of the renormalised light chiral condensate $\partial \Delta_{l, R}/\partial T$. For a vanishing baryon chemical potential, $\mu_B=0$, we obtain 
\begin{align} 
T_{c} =157.5(5)\,\textrm{MeV}\,.
\label{eq:TcMuB0}
\end{align} 
This is in quantitative agreement with the lattice results $156.5 \pm 1.5$ MeV  (HotQCD collaboration) \cite{HotQCD:2018pds}, $158 \pm 0.6$ MeV (Wuppertal-Budapest collaboration) \cite{Borsanyi:2020fev} and with the other state-of-the-art functional QCD results \cite{Fu:2019hdw, Gao:2020fbl, Gunkel:2021oya, Pawlowski:2025jpg}. 

In summary, the present functional QCD approach at strangeness neutrality, based on the fRG-approach with emergent composites used successfully at $\mu_S=0$, \cite{Fu:2019hdw, Ihssen:2024miv, Pawlowski:2025jpg}, passes all available benchmark tests at $\mu_B=0$. \\[-1.5ex]

\begin{figure}[t]
	\includegraphics[width=0.49\textwidth]{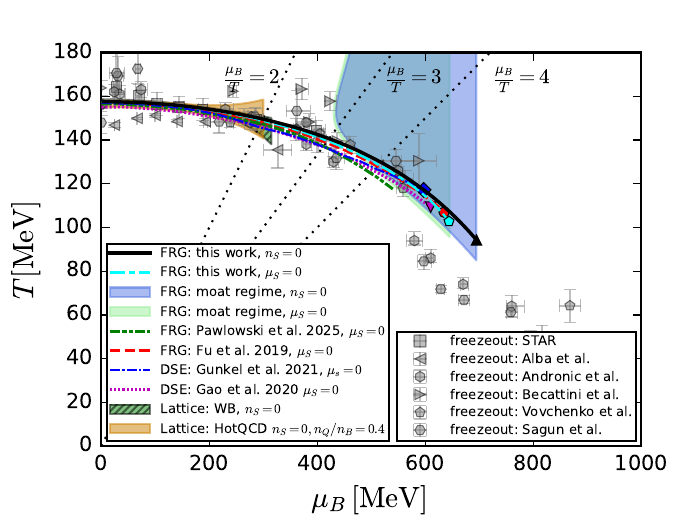}
	\caption{QCD phase diagram in comparison to the previous fRG \cite{Fu:2019hdw, Pawlowski:2025jpg}, DSE \cite{Gao:2020fbl, Gunkel:2021oya} and lattice QCD results \cite{HotQCD:2018pds, Borsanyi:2020fev}. Freeze-out points \cite{STAR:2017sal, Alba:2014eba, Andronic:2017pug, Becattini:2016xct, Vovchenko:2015idt, Sagun:2017eye} are also plotted for comparison. The moat regions are only plotted in the area where $\mu_B<{\mu_{B}}_{_\text{CEP}}$.
	}\label{fig:QCD_phasediagram}
\end{figure}

\emph{Phase structure of QCD with strangeness neutrality.--}
We are now in the position to map out the chiral phase structure of QCD at strangeness neutrality. For that purpose we have computed the crossover temperatures for $n_s=0$. Furthermore, we have also computed the crossover temperatures for $\mu_S=0$, which can be compared with \cite{Fu:2019hdw, Pawlowski:2025jpg}. 
Our results are shown in \Cref{fig:QCD_phasediagram} together with results for the QCD phase structure from functional QCD \cite{Fu:2019hdw, Pawlowski:2025jpg} and \cite{Gao:2020fbl} (DSE), where $\mu_S=0$ is implemented except in \cite{Gunkel:2021oya} one has a vanishing strange quark chemical potential $\mu_s=0$. We also show lattice results \cite{HotQCD:2018pds, Borsanyi:2020fev} at strangeness neutrality, $n_S = 0$. \Cref{fig:QCD_phasediagram} also contains the moat regimes for $n_S=0$ and $\mu_S=0$ computed in the present work. 

We start the discussion of the results in the low density regime, and specifically with the $\mu_B$-dependence of the crossover line. It has been already shown in \cite{Fischer:2014ata, Fu:2019hdw, Gao:2020fbl, Gunkel:2021oya, Pawlowski:2025jpg}, that the crossover line at $\mu_S=0$ is fit well even up to the ONP regime with a low order fit with two expansion coefficients $\kappa_2,\kappa_4$, 
\begin{align}
	\frac{T_c(\mu_B)}{T_c}=1-\kappa_2 \left(\frac{\mu_B}{T_c}\right)^2-\kappa_4 \left(\frac{\mu_B}{T_c}\right)^4+\cdots\,, 
	\label{eq:TcExpansion}
\end{align}
with $T_c=T_c(\mu_B=0)$. For a compilation of results for the curvature coefficients $\kappa_2,\kappa_4$ of various functional and lattice QCD computations for both $n_S=0$ and $\mu_S=0$ see \cite{Fischer:2026uni}. This also allows for a further benchmark test with lattice data. 

We have compiled selected results from functional QCD and lattice QCD in \Cref{tab:kappa}. For the comparison of lattice and functional QCD results one has to take into account the different computational procedures: on the lattice, $\kappa_2$ and $\kappa_4$ are computed as coefficients of a Taylor expansion in powers of $\mu^2_B/T_c^2$. In turn, in functional QCD including the present work the curvature coefficients are computed with a \textit{global} $\chi^2$-fit to the regime $\mu_B/T< 4$. There, the functional results are quantitatively reliable with a conservative 10\% systematic error estimate, for a detailed account of this error analysis see \cite{Fischer:2026uni}. We have also applied this procedure to the data of the recent work \cite{Pawlowski:2025jpg} for the sake of comparability, even though the quantitative reliability regime in \cite{Pawlowski:2025jpg} extends to $\mu_B/T \approx 4.5$. 

The functional QCD results for the curvature coefficient $\kappa_2$ agree very well with each other  within the systematic error estimate. The $\kappa_4$'s have a large numerical error due to its subleading nature as well as the coarseness of the temperature grid used for the computation of the crossover temperature: in most cases the condensates are resolved within $~1$\,MeV steps. Despite the global nature of the fits in functional QCD versus the definition as Taylor coefficients on the lattice, the $\kappa_2$'s are in excellent agreement within the respective errors. 

%
%

We close this discussion with a comparison of the ratio $\kappa_s(n_S=0)/\kappa_s(\mu_S=0)$. In general, such ratios of observables are less susceptible to systematic errors if the latter are global. Moreover, we expect this ratio to be smaller than unity.  Qualitatively this can already be inferred from the left panel of \Cref{fig:muS0} and \labelcref{eq:muB-muS}. The requirement of strangeness neutrality reduces $\mu_s$ for a given baryon chemical potential, $\mu_s(n_S=0)< \mu_s(\mu_S=0)$.  Consequently the finite density or finite chemical potential effects are delayed at strangeness neutrality, relative to $\mu_S=0$. We find 
\begin{align} 
	\frac{\kappa_2^{n_S=0}}{\kappa_2^{\mu_S=0}}=0.897(20) \,, \qquad \frac{\kappa_{2,\textrm{lat}}^{n_S=0}}{\kappa_{2,\textrm{lat}}^{\mu_S=0}}=0.893(35)\,. 
	\label{eq:Ratiokappa2s}
\end{align}
In \labelcref {eq:Ratiokappa2s} also contains the lattice result from \cite{Ding:2024sux}. We note that the functional and lattice QCD ratios agree on the percent level (0.8\%), while the curvature coefficients themselves show deviations of roughly 4\%. This corroborates the quantitative agreement between the lattice and functional QCD approaches at $\mu_B=0$ and provides further support for the functional QCD computation at larger density. 

%
\begin{table*}[t]
	\begin{center}
		\begin{tabular}{c|c c|c}
			\hline\hline & & &  \\
			& $\kappa_{2}$ &$\kappa_{4} $& $(T_{_\text{CEP}},{\mu_{B}}_{_\text{CEP}})$ [MeV] \\[2ex]
			\hline & & & \\[-2ex]
			functional QCD (this work, $n_{S}=0$)             &\, 0.0130(2) \,&\, 0.00014(5) \,& (92, 696) \\[1ex]
			functional QCD (this work, $\mu_{S}=0$)             & 0.0145(2) & 0.0002(5) & (102, 644) \\[1ex]
			functional QCD (\cite{Fu:2019hdw}, $\mu_{S}=0$)             & 0.0142(2) & 0.00029(2) & (107, 635) \\[1ex]
			functional QCD (\cite{Gao:2020fbl}, $\mu_{S}=0$)             & 0.0147(5) & 0.00028(2) & (109, 610) \\[1ex]
			functional QCD (\cite{Gunkel:2021oya}, $\mu_{s}=0$)             & 0.0167(5) & 0.00005(5) & (117, 600) \\[1ex]
			functional QCD (\cite{Pawlowski:2025jpg}, $\mu_{S}=0$)             & 0.0148(4) & 0.00047(5) & potential instability  \\[1ex]
			\hline  & & & \\[-2ex]
			lattice QCD (\cite{Ding:2024sux}, $n_{S}=0$)             & 0.0134(14) & ---  & ---  \\[1ex]
			lattice QCD (\cite{Ding:2024sux}, $\mu_{S}=0$)             & 0.015(1) & ---  & ---  \\[1ex]
			lattice QCD (\cite{Borsanyi:2020fev}, $n_{S}=0$)             & 0.0153(18) & 0.00032(67)  & ---  \\[1ex]
			\hline\hline
		\end{tabular}
		\caption{Curvature of the chiral phase boundary and location of CEP with different constraints.} 
		\label{tab:kappa}
	\end{center}\vspace{-0.5cm}
\end{table*}
%
%
\begin{figure}[b]
	\includegraphics[width=0.47\textwidth]{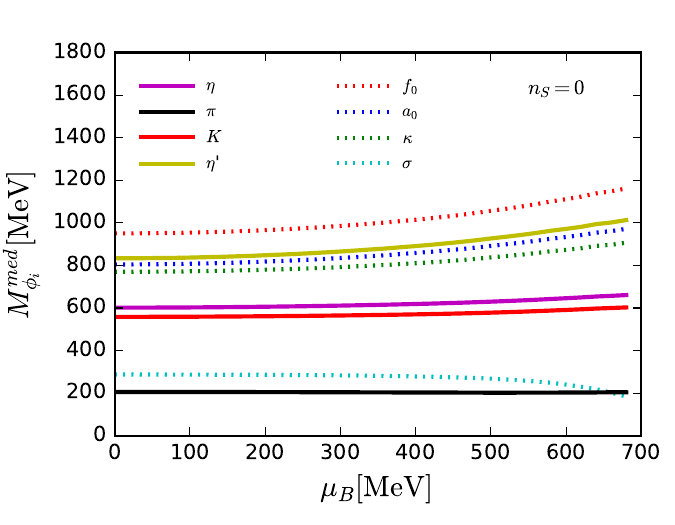}
	\caption{Meson masses as functions of the chemical potential at the pseudo-critical temperature at strangeness
		neutrality.}
		\label{fig:MesonMedium}
\end{figure}
%
We proceed with a discussion of the high density regime. At strangeness neutrality, $n_S=0$, the CEP is located at far larger baryon chemical potential, 
\begin{align}
	(T\,,\,{\mu_{B}})_{_\text{CEP}}=(92\,,\,696)\, \text{MeV}\,, 
	\label{eq:CEPns=0} 
\end{align}
see \Cref{fig:QCD_phasediagram}. Using the current setup at $\mu_S=0$ leads us to    $(T\,,\,{\mu_{B}})_{_\text{CEP}}=(102,644)\, \text{MeV}$. The latter location agrees well within a few percent with the result in \cite{Fu:2019hdw}, listed in \Cref{tab:kappa}. This is very remarkable as the location of the critical end point is very sensitive to small physics changes. It also suggests that the approximations are (still) quantitative in the regime $\mu_B/T\gtrsim 4$. 

In \Cref{tab:kappa} we have  compiled the predictions for the chiral critical end point of the functional QCD works shown in \Cref{fig:QCD_phasediagram}. We emphasise in this context that in the present approximation, and those used in \cite{Fu:2019hdw, Gao:2020fbl, Gunkel:2021oya}, the existence or absence of the CEP cannot be proven as the dominance of the scalar-pseudoscalar sector is assumed. However, the results in these works and the present one entail, that in the \textit{absence} of novel emergent dynamics at finite density, QCD with $\mu_S=0$ does not admit a chiral crossover beyond the CEP found in these works: the respective CEP interval is given by $\bigl(115-105\,,\, 600-650\bigr)\textrm{MeV}$ with a small width, see \cite{Fischer:2026uni}. This suggests a similar statement for QCD with strangeness neutrality with an interval of the same length roughly centred at 
\labelcref{eq:CEPns=0}. 

In any case, the large difference of the CEPs for $n_S=0$ and $\mu_S=0$ hints at a different dynamics in the large density regime for these cases. A natural candidate for this dynamics are additional soft modes, e.g.~soft kaons. This scenario can be checked  by following the masses of the scalar-pseudoscalar octet along the chiral crossover line. In  \Cref{fig:MesonMedium} we show the in-medium meson masses $M^\textrm{med}_{\phi^i}$  introduced in \cite{Fu:2019hdw}, see also the supplement. They give access to the masses of the (off-shell) mesons in the non-perturbative loops that govern the flow equation. Interestingly, the in-medium masses of the kaons even grow towards larger baryon chemical potential. Moreover, the pion in-medium mass is rather stable with a relatively large thermal part. The mass of the $\sigma$-mode is also rather stable and drops below the pion mass close to the CEP at roughly 650\,MeV. This behaviour is similar to that observed at $\mu_S=0$, for more details see the Supplement. Moreover, it has also been seen in \cite{Pawlowski:2025jpg}, where the full momentum dependence of pions and $\sigma$-mode has been resolved. However, with the deepening of the moat we indeed expect softer modes, which deserves further studies in particular for $\mu_B/T\gtrsim 4$. Such an analysis goes beyond the scope of the present work and will be provided elsewhere. 

We close this discussion with the interesting observation, that in contradistinction to the large change of the CEP location, the onset regime of the moats for $\mu_S=0$ and $n_S=0$ is roughly the same, see \Cref{fig:QCD_phasediagram}. The related comparison with the onset of the moat from \cite{Fu:2019hdw} as well as that in \cite{Pawlowski:2025jpg} has been deferred to  the supplement. There, we also show a Figure with all available data on the phase structure. We have refrained from showing all the data in \Cref{fig:QCD_phasediagram} for the sake of accessibility. \\[-1.5ex]

\emph{Conclusions and summary.--} 
We have computed the 2+1 flavour QCD phase structure at strangeness neutrality within functional QCD. We have used an approximation, in which the dominance of the chiral dynamics is assumed. This assumption has been checked rather comprehensively in functional QCD, and the only alternative scenario left is that of a regime with an instability, see \cite{Pawlowski:2025jpg} and the recent review \cite{Fischer:2026uni}. In turn, with the assumption, the QCD phase structure features a critical end point  at $(T\,,\,{\mu_{B}})_{_\text{CEP}}=(92\,,\,696)\, \text{MeV}$, see \labelcref{eq:CEPns=0}.  The present computation satisfies all QCD benchmarks at small baryon chemical potential. Specifically, the reduction of the curvature of the phase boundary in comparison to $\mu_S=0$ is in excellent agreement with the lattice results, see \labelcref{eq:Ratiokappa2s}. \\[-1.5ex]

\emph{Acknowledgements.--}
We thank the members of the fQCD collaboration \cite{fQCD} for discussions. This work is supported by the National Natural Science Foundation of China under Contract No.\ 12447102. This work is funded by the Deutsche Forschungsgemeinschaft (DFG, German Research Foundation) under Germany’s Excellence Strategy EXC 2181/1 - 390900948 (the Heidelberg STRUCTURES Excellence Cluster) and the Collaborative Research Centre SFB 1225 - 273811115 (ISOQUANT). Shi Yin is supported by the Alexander von Humboldt foundation.

\bibliography{ref-lib}

\newpage

\appendix 
\renewcommand{\thesubsection}{{S.\arabic{subsection}}}
\setcounter{section}{0}
\titleformat*{\section}{\centering \Large \bfseries}

\onecolumngrid

\section*{Supplemental Materials}

The supplemental materials detail our approximation to the full effective action of QCD in \Cref{app:ApproxEffAct}. The generalised flow equation for the effective action, the projection procedure onto the flow of the different couplings as well as the details of the fRG-approach with emergent composites are discussed in \Cref{app:flow_equation}. The determination of the fundamental ultraviolet parameters of QCD, the strong coupling and the current quark masses, and further details are provided in \Cref{app:set-up}. Finally, in and some further interesting results at vanishing and finite baryon chemical potential are compiled in \Cref{app:Results}. 

Before we go into the details, we would like to provide a brief overview and embedding of the present work in the functional QCD landscape. For a comprehensive overview over functional QCD and the different approximation schemes used in different works see \cite{Fischer:2026uni}. The works \cite{Fu:2019hdw, Ihssen:2024miv, Pawlowski:2025jpg} map out the phase structure of 2+1 flavour QCD at $\mu_S=0$, that is identical chemical potentials for all quarks, $\mu_s= \mu_l=\mu_B/3$, while \cite{Ihssen:2024miv} advances the approximation and the approach to systematic error estimates.  The approximation to the effective action if detailed in the Supplement, here we concentrate on the most important differences and advances in comparison to \cite{Fu:2019hdw, Fu:2024rto, Ihssen:2024miv, Pawlowski:2025jpg}.  

The previous works took into account the resonant scalar-pseudoscalar four-quark interaction channels of the pions $\boldsymbol{\pi}$ and $\sigma$-mode with emergent composites \cite{Gies:2001nw, Gies:2002hq, Pawlowski:2005xe, Fu:2019hdw, Fukushima:2021ctq}, for further applications and developments see \cite{Mitter:2014wpa, Braun:2014ata, Rennecke:2015eba, Cyrol:2017qkl}. In the emergent composite approach to QCD, higher order scatterings in the resonant channels are encoded in the effective potential $V(\rho)$ with $\rho= (\sigma^2 +\boldsymbol{\pi}^2)/2$. In \cite{Fu:2019hdw, Fu:2024rto} the potential was expanded in a Taylor expansion in $\rho$ around the solution of the mesonic equations of motion up to order 5 - 7, and the comprehensive analysis in \cite{Ihssen:2024miv} with an all order potential confirmed the quantitative nature of this expansion.

In the present work we extend these work by considering the full effective potential of the mesonic composite $\phi$. Together with the potential of the temporal background gluon we are led to 
\begin{align}
	V(\phi, A_0) = V_{A_0}(A_0) +V_\phi(\phi)\,. 
	\label{eq:FullPotential}
\end{align}
In \labelcref{eq:FullPotential} we have dropped the interaction terms between the temporal glue background and the nonet, and in the numerical application we resort to a converging Taylor expansion, building on the analysis of \cite{Ihssen:2024miv}. In summary, \labelcref{eq:FullPotential} and the respective mesonic propagators takes into account multi-scattering events of the scalar-pseudoscalar nonet channels. This includes potentially important contributions involving strange mesons, especially the additional pseudo-Goldstone bosons from the pseudoscalar octet, $K$ and $\eta$ mesons.

\subsection{Effective action and approximation scheme}
\label{app:ApproxEffAct}

In the present work we work use the fRG-approach to QCD, for QCD-related reviews see \cite{Litim:1998nf, Berges:2000ew, Schaefer:2004en, Pawlowski:2005xe, Gies:2006wv, Braun:2011pp, Dupuis:2020fhh, Fu:2022gou} and the recent review on the functional QCD approach to the phase structure of QCD \cite{Fischer:2026uni}.  

As already mentioned above, we use the fRG approach with emergent composites, where the effective action not only depend on the fundamental gluons, ghosts and quarks, but also emergent composite fields, in our case the scalar-pseudoscalar nonet. Moreover, we define the approximation scheme in terms of the rescaled field
\begin{align}
    \bar \Phi=\boldsymbol{Z}_{\Phi}^{1/2}\,\Phi,\qquad\text{with} \qquad \Phi=\{A,\bar c, c, \bar q, q, \phi\}\,,\quad \textrm{where}\quad \phi=(\sigma^a\,,\,\pi^a)^T\,,\quad a=0,...,8\,, 
    \label{eq:ScaledPhi}
\end{align}
where the $\sigma^a$ and $\pi^a$ are the scalar and pseudoscalar modes in the singlet-octet basis, and $\boldsymbol{Z}_{\Phi}=\textrm{diag}(Z_{\Phi_i})$ is the diagonal wave function matrix. The field $\bar\Phi$ in \labelcref{eq:ScaledPhi} is sometimes called the 'renormalised' field. However, it is indeed RG-invariant under the RG-transformations of the full theory at $k=0$, for a respective discussion see e.g.~\cite{Pawlowski:2005xe, Dupuis:2020fhh} and in particular \cite{Ihssen:2024miv}. The RG-invariance of the fields entails that all the dressings of $n$-point vertices are RG-invariant. On the present level it is simply a convenient choice which facilitates the book-keeping. However, the RG-invariant expansion scheme has better convergence property is fully momentum-dependent approximations are used, for a detailed discussion see \cite{Ihssen:2024miv}.  

Finally, we simplify the notation and relabel $\bar\Phi\to\Phi$ in a slight abuse of notation. Note that while the rescaled field is RG-invariant, it is not cutoff-independent, 
\begin{align}
    \partial_{t} \Phi=-\frac{1}{2}\eta_{\Phi}\Phi\,,\qquad \qquad \eta_{\Phi}=-\frac{1}{\boldsymbol{Z}_{\Phi}}\partial_{t}\boldsymbol{Z}_{\Phi}\,, 
   \label{eq:etaPhi}
\end{align}
with the diagonal matrix of anomalous dimensions $\eta_{\Phi}=\textrm{diag}(\eta_{\Phi_i})$ which accounts for the anomalous scaling with $k$.

\subsubsection{Effective action}
\label{app:EffAct}

We use the following approximation of the effective action $\Gamma_k$ for $N_f=2+1$ flavour QCD, 
\begin{subequations} 
\label{eq:Approximation+Notation}
\begin{align}\nonumber 
	\Gamma_k=&\,\int\limits_0^\beta \!\! d x_0\int\!\!  d^3 x\, \Biggr \{ \frac{1}{4} F_{\mu\nu}^a F_{\mu\nu}^a+ (\partial_\mu \bar c^a) D_{\mu}^{a b} c^{b}+\frac{1}{2\xi}(\partial_\mu A_{\mu}^{a})^2 +\frac{1}{2} A^a_{\mu}(-p) \Big({\Gamma_{AA}^{(2)}}_{\mu \nu}^{ab}- \Pi_{\mu \nu}^{\perp}\delta^{ab} p^2 \Big)A _{\nu}^b(p)+V_{A_0}(A_0)\\[2ex]\nonumber
	&+\bar q\,  \Big[\gamma_\mu D_\mu-\gamma_0 \hat \mu_q+  h_q^{1/2} \Sigma_{5}  \cdot  h_q^{1/2} \Big]\,q  
    - \sum_{a=0}^8 \left( \lambda_{q_a} \left[ \bar q\, T^a \,q\right]^2+\lambda_{q_a} \left[ \bar q\, i \gamma _5 T^a \,q\right]^2 \right) \\[2ex]
	&+\tr \left[ \bar{D}_\mu\,\Sigma   \, \bar{D}_\mu \,\Sigma^\dagger \right] 
 +V_\phi(\Sigma,\Sigma^\dagger) \Biggl \} \,,
	\label{eq:Effective_action}
\end{align}
with  $\beta = 1/T$ and the rescaled gluon fields. The transversal projector $\Pi_{\mu\nu}^{\perp}(p)=\delta_{\mu\nu}-p_{\mu}p_{\nu}/p^2$. The rescaled quark and meson fields are given by  
\begin{align}
q=(u,d,s)^T\,,\qquad \qquad  \Sigma=T^a(\sigma^a+i \pi^a)\,, \qquad 	\Sigma_5 =T^a(\sigma^a+i \gamma_5 \pi^a)\,,\qquad  a=0,1,...,8\,,
\label{eq:Quarks+Mesons}
\end{align}
where $T^a=\lambda^a/2$ with the Gell-Mann matrices $\lambda^a$ for $a=1,...,8)$ are the generators of the flavour su(3)-algebra and $T^{0}=\mathbb{1}_{N_{f}\times N_{f}}/\sqrt{2N_{f}}$. The $\sigma^a$ and $\pi^a$ carry the quantum numbers of the respective mesons and feature the same pole masses. However, these \textit{emergent composites} are introduced as efficient book-keeping devices for resonant four-quark interaction channels and should not be confused with on-shell asymptotic particles.  Furthermore, their occurrence in \labelcref{eq:Effective_action} does not signal an low energy effective theory, it can be shown that the QCD effective action in the fRG-approach with emergent composites reduces to the standard effective action on the equations of motion for the composite fields, similarly to the better-known two-particle irreducible (2PI) approach, where the 2PI effective action reduces to the 1PI effective action on the gap equation (equation of motion for the propagator field). 

The quark chemical potential $\hat \mu_q$ and the Yukawa coupling $h_q$ in \labelcref{eq:Effective_action} are taken diagonal, 
\begin{align} 
\hat\mu_q = \begin{pmatrix}
\mu_l&0&0\\
0&\mu_{l}&0\\
0&0&\mu_s \end{pmatrix}\,,\qquad \qquad\qquad  h_q=\begin{pmatrix}
  h_{l}&0&0\\
0&  h_{l}&0\\
0&0&  h_{s}
\end{pmatrix}\,, 
\label{eq:hatmu+Yukawa}
\end{align}
\end{subequations}
and we have identified the up and down chemical potentials and Yukawa couplings. The quark chemical potentials $\mu_{q_i}$ can be expressed in terms of the baryon chemical potential, that of the electric charge and strangeness, 
\begin{align}
	\mu_u = \frac{1}{3}	\mu_B + \frac{2}{3} \mu_Q\,, \qquad 
	\mu_d = \frac{1}{3}	\mu_B - \frac{1}{3} \mu_Q \,,\qquad 
	\mu_s = \frac{1}{3}	\mu_B - \frac{1}{3} \mu_Q - \mu_S\,.
\label{eq:DifferentMus}
\end{align}
Hence, the identical light quark chemical potentials $\mu_u=\mu_d=\mu_l$ entail $\mu_Q=0$ and $\mu_s= \mu_l-\mu_S$.  All couplings and wave functions in \labelcref{eq:Approximation+Notation} depend on the momentum cutoff scale $k$ and this dependence is computed from the flow equation of the effective action discussed in \Cref{app:flow_equation}. 

The first line on the right-hand side of \labelcref{eq:Effective_action} comprises the approximation of the pure glue sector. In this work, we adopt the same truncation for the glue sector as that in \cite{Fu:2019hdw, Fu:2024rto, Ihssen:2024miv, Pawlowski:2025jpg}. The quantitative nature of this approximation has been tested comprehensively in \cite{Fu:2019hdw, Ihssen:2024miv} against fRG benchmark results \cite{Cyrol:2017qkl, Fu:2025hcm} in the vacuum with full momentum dependence as well as lattice results (gluon propagator) at finite temperature. Further details will be provided in \Cref{app:GlueSector}. 

The two other lines comprise the approximation of the matter sector, further details will be provided in  \Cref{app:Matter_sector} and \Cref{app:mesonmass}, and we only provide a brief overview here: We only consider the baryon number and strangeness chemical potentials and ignore the isospin chemical potential, see \labelcref{eq:hatmu+Yukawa}. This is a good approximation in the context of heavy-ion collisions, as charge conservation leads to only very small isospin chemical potentials \cite{Fu:2018qsk, Fu:2018swz, Wen:2019ruz}. $\Sigma$ denotes the nonets of scalar and pseudoscalar meson fields, and the meson effective potential $\tilde V_\Sigma (\Sigma,\Sigma^\dagger)$ including chiral symmetric part, the chiral symmetry breaking term (linear sigma terms) and the $U_A(1)$ symmetry breaking term (the Kobayashi-Maskawa-'t~Hooft term \cite{Kobayashi:1970ji, tHooft:1976rip, tHooft:1976snw}), see \Cref{app:Matter_sector} for more details. The meson curvature masses are obtained from the eigenvalues of the Hessian matrix of the meson effective potential. The meson fields are coupled to the quarks through the Yukawa interactions. In this work, we distinguish the light and strange Yukawa couplings and the corresponding dynamical hadronisation, see \Cref{app:flow_equation}. \\[-1.5ex]

\subsubsection{Matter sector of $N_f=2+1$ flavour QCD}
\label{app:Matter_sector}

The nonets of scalar and pseudoscalar meson fields have been defined in \labelcref{eq:Quarks+Mesons}. The covariant derivative acting on the meson fields reads 
\begin{align}
\bar{D}_\mu\Sigma=\partial_\mu \Sigma + [\delta_{\mu 0} \hat{\mu},\Sigma]\,.
\end{align}
In this work, we work in the isospin-symmetric approximation. Then, there are two non-vanishing vacuum expectation values, i.e.~the light and strange condensates $\sigma_l$, $\sigma_s$. In \labelcref{eq:Effective_action}, the meson fields are coupled to the quarks through the scalar-pseudoscalar symmetric Yukawa term and the Yukawa coupling matrix is a diagonal matrix in the flavour space, see \labelcref{eq:hatmu+Yukawa}. 

The meson effective potential can be divided into three parts based on the symmetry:
\begin{align}
  {V}_{\Sigma}(\Sigma,\Sigma^\dagger)&=V_{\chi}(\rho_1,\rho_2,\rho_3)-c_A \xi\,, -c_l\sigma_l-c_s\sigma_s/\sqrt{2}
  \label{eq:tildeU}
\end{align}
where 
\begin{align}
	\xi=\det(\Sigma)+\det(\Sigma^\dagger)\,.
\label{eq:thooftDet}
\end{align}
In  \labelcref{eq:tildeU}, $V_{\chi}(\rho_1,\rho_2,\rho_3)$ is the chirally symmetric part of the meson effective potential and has the full chiral $U_L(3)\times U_R(3)$ symmetry. In general, it depends on three chiral invariants and we choose 
\begin{align}
	\rho_1=\text{tr}(\Sigma \,\Sigma ^\dagger) \,, \qquad \rho_2=\sqrt{6\, \text{tr}\bigg (\Sigma \, \Sigma ^\dagger-\frac{1}{3} \rho_1 \mathbb{1}_{3 \times 3} \bigg )^2 } \,,\qquad \rho_3 \propto \text{tr}(\Sigma \, \Sigma ^\dagger)^3\,.
	\label{eq:rho123} 
\end{align}
For more details see e.g.~\cite{Jungnickel:1996aa, Schaefer:2008hk, Mitter:2013fxa, Herbst:2013ufa, Rennecke:2016tkm, Wen:2018nkn, Fu:2018qsk, Fu:2019hdw, Pawlowski:2025jpg}. For vanishing pseudoscalar field $\pi^a=0$, the invariants $\rho_1,\rho_2$ have simple forms in terms of the scalar fields $\sigma_l,\sigma_s$. The latter fields are combinations of the fields $\sigma^0, \sigma^8$ in the singlet-octet basis with 
\begin{align}
	\begin{pmatrix} \sigma^{\ }_l \\[.5ex] \sigma^{\ }_s \end{pmatrix}  =
	\frac{1}{\sqrt{3}} \begin{pmatrix} \sqrt{2} &  1
		\\[.5ex]  1 & -\sqrt{2} \end{pmatrix}\,
	\begin{pmatrix} \sigma^{\ }_0 \\[.5ex]  \sigma^{\ }_8 \end{pmatrix} \,, 
	\label{eq:SigmaSigmas}
\end{align}
see again~\cite{Jungnickel:1996aa, Schaefer:2008hk, Mitter:2013fxa, Herbst:2013ufa, Rennecke:2016tkm, Wen:2018nkn, Fu:2018qsk, Fu:2019hdw}. With \labelcref{eq:SigmaSigmas} we find 
\begin{align}
	\rho_1=\frac{1}{2}(\sigma_l^2+\sigma_s^2) \, , \qquad \qquad  \rho_2=\frac{1}{2}(2 \sigma_s^2-\sigma_l^2)\,.
\label{eq:rho12} 
\end{align}
In the following, we shall drop the invariant $\rho_3$ for simplicity, see e.g.~\cite{Mitter:2013fxa, Rennecke:2016tkm}. This is supported by its higher order nature as well as the dominance of the contributions from the quark loops, were we follow the argument in \cite{Ihssen:2024miv}: The light and strange quark loops generate functions $V_{q_l}(\rho_l)$ and $V_{q_s}(\rho_s)$ respectively. Only the induced field-dependence of the Yukawa couplings mixes the invariants and these terms are subleading. For this reason we have dropped them in the present work to begin with. Further mixings are triggered by the meson contributions to the effective potential, which are subleading in a large regime in the phase structure. They become even dominant close to a critical point due to their coupling to the soft modes. For the chiral limit these are the pions, see \cite{Braun:2023qak}, and for the potential critical end point this is a mixture of the $\sigma$-mode, the density mode and the Polyakov loop, see \cite{Haensch:2023sig}. However, in this regime the subleading nature follows from the the efficient scale hierarchy in QCD and the \LEGO-principle. The letter entails, that relevance arguments carry over from subsectors of QCD to QCD, if the (quark-gluon) interfaces are well-controlled, see \cite{Ihssen:2024miv, Fischer:2026uni}. In summary, this allows us to only consider the invariants $\rho_{1,2}$ in \labelcref{eq:rho123}. 

The linear breaking terms $c_l \sigma_l+c_s\sigma_s$ can be understood as source terms for the zero modes (constant fields) $\sigma,\sigma_s$ and are cutoff independent, see e.g.~\cite{Fu:2019hdw}. They account for the explicit chiral symmetry breaking that comes with the current quark masses. As we have parametrised  \labelcref{eq:tildeU} in terms of the rescaled fields, the coefficients $c_{l,s,A}$ satisfy 
the flow equations 
\begin{align} 
\partial_t \left( c_l \sigma_l+c_s\sigma_s\right) =0 \,,\qquad \longrightarrow\qquad \partial_t c_i = \frac12 \eta_i\,c_i\,,\qquad i=l,s\,.
\label{eq:Flowc_i}
\end{align}
The $k$- and $\rho_i$-dependence of $V_\chi$ is determined by its flow equation, see \Cref{app:MesonSector}. In the present work we consider a Taylor expansion in $\rho_1,\rho_2$ around the solutions $\kappa_1=\rho_l^\textrm{\tiny{EoM}}$ and $\kappa_2= \rho_s^\textrm{\tiny{EoM}}$ of the mesonic equations of motion, 
\begin{align} 
	\frac{\partial ( V_\chi- c_A\, \xi)}{\partial \sigma_{l}}(\sigma^\textrm{\tiny{EoM}}_l, \sigma^\textrm{\tiny{EoM}}_s)=c_{l}\,, \qquad \quad \frac{\partial  ( V_\chi- c_A\, \xi)}{\partial \sigma_{s}}(\sigma^\textrm{\tiny{EoM}}_l, \sigma^\textrm{\tiny{EoM}}_s)=c_{s}\,, \qquad \phi^\textrm{\tiny{EoM}}=(\sigma_0^\textrm{\tiny{EoM}}, 0,...,\sigma_8^\textrm{\tiny{EoM}}, 0,...,0)^T\,.
	\label{eq:sigmaEoM}
\end{align} 
This leads us to 
\begin{align}
V_{\chi,k} (\rho_1,\rho_2)=\sum_{i,j=0}^{i+j\leq N} \frac{\lambda_{ij,k}}{i!j!}(\rho_1 -\kappa_{1,k})^i(\rho_2-\kappa_{2,k})^j\,,
\label{eq:TaylorVchi} 
\end{align}
with the rescaled fields $\rho_i$. Hence, the expansion coefficients $\lambda_{ij,k}$ are RG-invariant and can be interpreted as the scattering strength of the resonant scattering of $2 i +2 j$ fermionic bilinears. They are determined by the coupled set of flow equations for the full QCD system and we monitor the results with an increasing order of $N$ with $i+j\leq N$. We find convergence with $N=5$. The $\kappa_{1,k}$, $\kappa_{2,k}$ are the RG scale $k$ dependent expansion points, which are always at the minimum of the effective potential for every value of $k$. 

We note in passing that we could have chosen expansion points different to $\kappa=(\kappa_1,\kappa_2)$. Indeed, it has been shown that an expansion about the \textit{flowing} solution $\kappa$ of the equations of motion is converging slower than a \textit{fixed} expansion about $\kappa_{k=0}$, see \cite{Pawlowski:2014zaa}. However, the latter procedure requires iterative runs of the full system and we take the more convenient choice \labelcref{eq:TaylorVchi}. We note that close to the critical end point a fixed expansion works qualitatively better as has been shown in \cite{Pawlowski:2014zaa}. However, for a quantitative resolution of this regime we will resort to computing the full potential without an expansion, see \cite{Pawlowski:2025jpg}. This goes beyond the scope of the present work and will be presented elsewhere.  

Notably, the choices of the expansion arguments $\rho_1, \rho_2$ are not unique, and the physical observables should not depend on the choices. However, it would simply the mathematical expressions of meson masses and improve the stability of numerical calculations by choosing a suitable set of expansion arguments for the effective potential. We have checked that our expansion scenario converges quickly and are consistent with the results obtained from other expansion arguments in \cite{Rennecke:2016tkm, Mitter:2013fxa}.

Finally, with \labelcref{eq:thooftDet}, $c_A \,\xi$ in \labelcref{eq:tildeU} is nothing but the Kobayashi-Maskawa-'t Hooft term \cite{Kobayashi:1970ji, tHooft:1976rip, tHooft:1976snw}. Since $\xi$ is a rescaled field, $c_A$ is the renormalisation-group invariant 't Hooft coupling. This term breaks $U_A(1)$ symmetry and is related to the mass splitting of $\eta$ and $\eta'$ meson. In this work, we introduce an instanton-induced 't Hooft coupling \cite{Pawlowski:1996ch} for $k \gtrsim 700$ MeV 
\begin{align}
 c_A \propto \exp\left(- \frac{2 \pi}{\alpha_{k}} \right) k^{-3 N_f+4}\,,
\end{align}
where $\alpha_{k}$ is the strong coupling. For $k \lesssim 700$ MeV it tends to a constant, see \Cref{app:set-up}.

\subsubsection{Meson and quark masses and scatterings}
\label{app:mesonmass}

In the fRG-approach with emergent composites the field-dependent light and strange quark mass functions are given by 
\begin{align}
	m_l(\phi)=\frac{h_{l}}{2}\sigma_l , \qquad m_s(\phi)=\frac{h_{s}}{\sqrt{2}}\sigma_s\,, 
\label{quarkmass_eq}
\end{align}
with field-dependent Yukawa couplings. The masses $m_{l},m_s$ in \Cref{quarkmass_eq} are RG-invariant, and the constituent quark masses are obtained by evaluating \labelcref{quarkmass_eq} on the equations of motion $\sigma^\textrm{\tiny{EoM}}_l, \sigma^\textrm{\tiny{EoM}}_s$ in \labelcref{eq:sigmaEoM}. 

The field-dependent meson mass functions are obtained from the Hessian of the meson effective potential, 
\begin{align}
	H_{ij}(\phi)=\frac{ \partial^2 {V}_{\phi} (\phi)}{\partial \phi_i \partial \phi_j} =\frac{ \partial^2\left(  {V}_{\chi}(\rho_1,\rho_2)-c_A\, \xi\right)}{\partial \phi_i \partial \phi_j}   \,.
\label{mesonmass_eq}
\end{align}
The detailed expressions can be found in \Cref{app:mesonmass} or e.g.~in \cite{Rennecke:2016tkm, Wen:2018nkn}. The physical masses are obtained by evaluating the Hessian on the mesonic EoM \labelcref{eq:sigmaEoM}. There, $\Sigma, \Sigma^\dagger$ derivatives can be converted into $\rho_{1,2}$-derivatives and the Hessian \labelcref{mesonmass_eq} can be written in terms of 
\begin{align} 
	V_\chi^{(n,m)}(\rho_1,\rho_2) := \frac{\partial^{(n+m)} V_\chi  (\rho_1,\rho_2)}{(\partial \rho_1^n \partial \rho_2 ^m)}\,, 
\end{align} 
For the sake of readability we also introduce the notation 
\begin{align} 
H_{s,ij} =H_{\sigma_{i}\sigma_{j}}\,,\qquad  \textrm{and}\qquad 	H_{p,ij} =H_{\pi_{i}\pi_{j}}\,, \qquad \textrm{with}\qquad i,j=l,s\,.
\label{eq:Hsp}
\end{align} 
With \labelcref{eq:Hsp} and the parametrisation \labelcref{eq:TaylorVchi}, the Hessian for the scalar sector on the mesonic equation of motion reads 
\begin{align}\nonumber 
	H_{s,ll}=&-\frac{c_A \sigma_s}{\sqrt{2}}+V_\chi^{(1,0)}-V_\chi^{(0,1)} + (V_\chi^{(2,0)} -2V_\chi^{(1,1)}+V_\chi^{(0,2)})\sigma_l^2\,, \\[1ex]\nonumber 
	H_{s,ls}=&-\frac{c_A \sigma_l}{\sqrt{2}}+ (V_\chi^{(2,0)}+V_\chi^{(1,1)}-2V_\chi^{(0,2)}) \sigma_l \sigma_s\,, \\[1ex]\nonumber 
	H_{s,ss}=&V_\chi^{(1,0)}+2 V_\chi^{(0,1)} +(V_\chi^{(2,0)}+4V_\chi^{(1,1)}+4V_\chi^{(0,2)})\sigma_s^2\,,  \\[1ex]\nonumber 
	H_{s,11}=&\frac{c_A \sigma_s}{\sqrt{2}}+V_\chi^{(1,0)} +\frac{7 \sigma_l^2 - 2 \sigma_s^2}{2 \sigma_s^2-\sigma_l^2}V_\chi^{(0,1)}\,, \\
	H_{s,44}=&\frac{c_A \sigma_l}{2}+V_\chi^{(1,0)}+\frac{\sigma_l^2+3 \sqrt{2} \sigma_l\sigma_s+4 \sigma_s^2}{2 \sigma_s^2-\sigma_l^2}V_\chi^{(0,1)}\,, 
\end{align}
where it is understood that $\sigma_{l,s}\to \sigma_{l,s}^\textrm{\tiny{EoM}}$ and 
\begin{align} 
	H_{ij}= \left. H_{ij}(\phi)\right|_{\phi=\phi^\textrm{\tiny{EoM}}}\,,\qquad \qquad V_\chi^{(n,m)}=V_\chi^{(n,m)}(\kappa_1,\kappa_2)\,.
\end{align} 
For the pseudoscalar sector we find 
\begin{align}\nonumber 
	H_{p,ll}=&\frac{c_A \sigma_s}{\sqrt{2}}+V_\chi^{(1,0)}-V_\chi^{(0,1)}\,,\\[1ex]\nonumber 
	H_{p,ls}=&\frac{c_A \sigma_s}{\sqrt{2}}\,,\\[1ex]\nonumber 
	H_{p,ss}=&V_\chi^{(1,0)}+2 V_\chi^{(0,1)}\,,\\[1ex]\nonumber 
	H_{p,11}=&-\frac{c_A \sigma_s}{\sqrt{2}}+V_\chi^{(1,0)}-V_\chi^{(0,1)}\,,\\[1ex]
	H_{p,44}=&- \frac{c_A \sigma_l}{2} + V_\chi^{(1,0)} + \frac{\sigma_l^2- 3 \sqrt{2} \sigma_l \sigma_s+4 \sigma_s^2}{2 \sigma_s^2-\sigma_l^2} V_\chi^{(0,1)}\,. 
	\label{eq:Hp}
\end{align}
The non-vanishing off-diagonal element $H_{s/p, ls}$ introduce mixing angles between the light-strange basis and physical basis, 
\begin{align}
\begin{pmatrix}f_0 \\[1ex] \sigma \end{pmatrix} 
= \begin{pmatrix} \cos\varphi_s & - \sin\varphi_s \\[1ex] \sin\varphi_s & \cos\varphi_s \end{pmatrix} \begin{pmatrix}\sigma_l \\[1ex] \sigma_s \end{pmatrix}\,,\qquad \qquad 
\begin{pmatrix}\eta \\[1ex] 
\eta^\prime \end{pmatrix} 
= \begin{pmatrix} \cos\varphi_p & - \sin\varphi_p \\[1ex]\sin\varphi_p & \cos\varphi_p \end{pmatrix} \begin{pmatrix}\eta_l \\[1ex] \eta_s \end{pmatrix}\,,
\label{eq:plstrafo}
\end{align}
with
\begin{align}
\varphi_{s/p}=\frac{1}{2} \arctan\Biggl[\frac{2H_{s/p,ls}}{H_{s/p,ss}-H_{s/p,ll}}\Biggr]\,.
\label{eq:mix_phi}
\end{align}
Then, the square of meson masses are given by
\begin{align}\nonumber
m_{f_0}^2&=\cos^2\varphi_s H_{s,ll}+\sin^2 \varphi_s H_{s,ss}-2 \sin \varphi_s \cos \varphi_s H_{s,ls}\,,\\[.5ex]\nonumber 
m_{\sigma}^2&=\sin^2\varphi_s H_{s,ll}+\cos^2 \varphi_s H_{s,ss}+2 \sin \varphi_s \cos \varphi_s H_{s,ls}\,,\\[.5ex]\nonumber 
m_{a_0}^2&=H_{s,11},\\[.5ex]\nonumber 
m_{\kappa}^2&=H_{s,44}\,,\\[.5ex]\nonumber 
m_{\eta}^2&=\cos^2\varphi_p H_{p,ll}+\sin^2 \varphi_p H_{p,ss}-2 \sin \varphi_p \cos \varphi_p H_{p,ls}\,,\\[.5ex]\nonumber 
m_{\eta'}^2&=\sin^2\varphi_p H_{p,ll}+\cos^2 \varphi_p H_{p,ss}+2 \sin \varphi_p \cos \varphi_p H_{p,ls}\,,\\[.5ex]\nonumber 
m_{\pi}^2&=H_{p,11}\,,\\[.5ex]
m_{K}^2&=H_{p,44}\,.
\label{eq:AllMesonMasses}
\end{align}
The masses of $f_0, \sigma, \eta,\eta'$ can be expressed concisely as  
\begin{align}\nonumber 
m_{f_0/\eta'}^2=&\, \frac12 \Biggl[ H_{s/p,ll}+H_{s/p,ll}+\sqrt{(H_{s/p,ll}-H_{s/p,ss})^2+4 H_{s/p,ls}^2}\,\, \Biggr]\,,\\[2ex]
m_{\sigma/\eta}^2=&\, \frac12 \Biggl[ H_{s/p,ll}+H_{s/p,ss}-\sqrt{(H_{s/p,ll}-H_{s/p,ss})^2+4 H_{s/p,ls}^2}\,\, \Biggr]\,.
\end{align}
The momentum-independent mesonic $n$-point correlation functions are also encoded in the effective potential. Here, we list the non-vanishing 3-point correlation functions used in the computation of $Z_{\pi^+}$ or rather that of  $\eta_{\pi^+}$, 
\begin{align}\nonumber 
V_{\pi^{+} \pi^{-} f_0}=&\, \cos\varphi_s\, \Big[ (V_\chi^{(0,2)} - 2 V_\chi^{(1,1)} + V_\chi^{(2,0)}) \sigma_l \Big]- \sin\varphi_s \,\Big[-(\frac{c_A}{\sqrt{2}} + (-2 V_\chi^{(0,2)} + V_\chi^{(1,1)} + V_\chi^{(2,0)}) \sigma_s \Big] \,,\\[2ex]\nonumber 
V_{\pi^{+} \pi^{-} \sigma}=&\,  \sin\varphi_s \,\Big[ (V_\chi^{(0,2)} - 2 V_\chi^{(1,1)} + V_\chi^{(2,0)}) \sigma_l \Big] + \cos\varphi_s \, \Big[-(\frac{c_A}{\sqrt{2}} + (-2 V_\chi^{(0,2)} + V_\chi^{(1,1)} + V_\chi^{(2,0)}) \sigma_s \Big] \,,\\[2ex]\nonumber 
V_{\pi^{+} a_0^{-} \eta}=&\,  -\cos\varphi_p\frac{6 V_\chi^{(0,1)} \sigma_l}{\sigma_l^2 - 2 \sigma_s^2}+ \sin\varphi_p \frac{c_A}{\sqrt{2}}\,,\\[2ex]\nonumber 
V_{\pi^{+} a_0^{-} \eta'}= & \, -\sin\varphi_p\frac{6 V_\chi^{(0,1)} \sigma_l}{\sigma_l^2 - 2 \sigma_s^2} - \cos\varphi_p \frac{c_A}{\sqrt{2}}\,,\\[1ex]
V_{\pi^{+} \kappa^{-} K^0} = &\,  V_{\pi^{+}K^{-}\kappa^0} = \frac{c_A}{\sqrt{2}} - \frac{6 V_\chi^{(0,1)} \sigma_s}{\sigma_l^2 - 2 \sigma_s^2}\,.
\end{align}
This concludes our discussion of the meson sector.

\subsection{Functional RG approach to QCD with emergent composites}
\label{app:flow_equation}

In this supplement we discuss the flow equations for the different parts of the effective action \labelcref{eq:Approximation+Notation}. We use the fRG-approach with emergent composites put forward in \cite{Gies:2001nw, Pawlowski:2005xe, Floerchinger:2009uf, Fu:2019hdw, Fukushima:2021ctq}. As mentioned in the introduction of \Cref{app:ApproxEffAct}, in vacuum QCD it has been applied to the scalar-pseudoscalar four-quark interaction channels of the pions  $\boldsymbol{\pi}$ and $\sigma$-mode, \cite{Gies:2002hq, Braun:2009ewx, Mitter:2014wpa, Braun:2014ata, Rennecke:2015eba, Cyrol:2017qkl, Ihssen:2024miv} and the vector channel has been considered additionally in \cite{Rennecke:2015eba}. Emergent composites in the full scalar-pseudoscalar channel have been used in many flavour QCD in the chiral limit in \cite{Goertz:2024dnz}. These different applications and the existence of precision computations in the vacuum, \cite{Mitter:2014wpa, Cyrol:2017qkl, Fu:2025hcm} allows for a comprehensive assessment of the systematic error of the approximation used here. 

The fRG approach with emergent composites has been applied to functional QCD at finite temperature and density in \cite{Fu:2019hdw, Fu:2024rto, Pawlowski:2025jpg}, using the generalised flow equation \cite{Pawlowski:2005xe}, for further developments see \cite{Fu:2019hdw, Fukushima:2021ctq, Ihssen:2023nqd}, 
\begin{subequations}
	\label{eq:GenFlow}
\begin{align}
		\partial_t \Gamma_k [\Phi]  + \dot{\Phi}_a \left(\frac{\delta \Gamma_k[\Phi]}{\delta \Phi_a} + c_{\sigma_i} \delta_{a\mathbf{\sigma}_i} \right) =\textrm{Flow}[\Phi] \,, \qquad \textrm{with}\qquad \textrm{Flow}[\Phi]= \frac{1}{2} G_{ac} [ \Phi]\,\left(\partial_t\delta^c_b + \frac{\delta \dot{\Phi}_b}{\delta  \Phi_c}\right) R^{ab} \,,
		\label{eq:GenFlowEq}
\end{align}
with $i=l,s$ and $(\sigma_i) = (\sigma_l,\sigma_s)$, the diagrammatic part  $\textrm{Flow}[\Phi]$ and the flowing field $\dot\Phi$. The flowing field induces a differential reparametrisation of the effective action at each flow step. It will be used to completely absorb the scalar-pseudoscalar four-quark scattering channels into the emergent composite $\phi$ and its dynamics. The zero mode term proportional to $c_{\sigma_j} $ on the left hand side of  \labelcref{eq:GenFlowEq} guarantees manifest chiral symmetry, see  \cite{Fu:2019hdw}, for further conceptual advances see \cite{Ihssen:2024ihp, Ihssen:2025cff, Ihssen:2025hyl}. 

The parameter $t= \ln (k/\Lambda_\textrm{UV})$ in \labelcref{eq:GenFlowEq} is the RG-time and the propagator in field space is given by
\begin{align}
	G_k [\Phi] = \frac{1}{\Gamma_k^{(2)}[\Phi] + R_k } \,, \qquad \textrm{with}\qquad G_{ab} = (G_k [\Phi])^{\ }_{\Phi_a \Phi_b} \,.
	\label{eq:PropPhi}
\end{align}
\end{subequations}
In \labelcref{eq:GenFlow} we have used DeWitt's condensed notation for a concise form of the rather long explicit impressions, see also \cite{Ihssen:2024miv}. The super-indices $a,b,c,...$ now also include space-time or momentum arguments and a contraction of indices also accommodates the respective integrations or Matsubara sums at finite temperature.  

For the explicit computation we use RG-adapted regulators, \cite{Pawlowski:2005xe}. 
They have the structure 
\begin{subequations} 
	\label{eq:Regs}
\begin{align} 
R_{\Phi_i} =Z_{\phi_i} {\cal T}_{\Phi_i}(p) r_{\Phi_i}(x)\,, 
\label{eq:RegPhii}
\end{align} 
with the dimensionless shape functions $r_{\phi_i}$. The prefactors ${\cal T}_{\Phi_i}(p)$ account for the dispersion of the field $\Phi_i$,  
\begin{align} 
	{\cal T}^{ab}_{A}(p)=p^2 \,\Pi^\bot\delta^{ab}\,,\qquad  {\cal T}_{c}(p)=p^2 \, \delta^{ab}\,, \qquad  {\cal T}^{AB}_{q_i}(p)=\slashed{p}  \,\delta^{AB} \,,\qquad {\cal T}_{\phi_i}(p)=p^2 \,, \qquad a=1,..,8\,,\quad A=1,2,3\,.
	\label{eq:Dispersions}
\end{align} 
The wave function factors $Z_{\phi_i}$ arrange for the RG-adaptation  \cite{Pawlowski:2005xe}: The RG-scaling (not $k$-scaling) of the regulator is that of a two-point function, leading to an RG-invariant (generalised) flow equation \labelcref{eq:GenFlow} for the effective action. This property leads to a better convergence of approximations and in particular a better convergence of critical scaling properties. In the present work we use $3d$ flat or Litim regulators, \cite{Litim:2001up} with the shape functions 
\begin{align} 
	r_{\Phi_i\neq q}(x)=  r_B(x)=\left( 1 -\frac{1}{x}\right)\,\theta(1 -x) \,,\qquad r_{q_i}(x)=r_F(x)=\left( 1 -\frac{1}{\sqrt{x}}\right)\,\theta(1 -x)\,,\qquad\qquad  x=\frac{\boldsymbol{p}^2}{k^2} \,.
\label{eq:ShapeFunctions} 
\end{align}
\end{subequations} 
From the generalised flow equation \labelcref{eq:GenFlow} for the effective action we infer the flow equations for the glue sector in   \Cref{app:GlueSector}, the flow of the effective potential $V_\phi$ in  \Cref{app:MesonSector}), and the anomalous dimensions of quarks, \Cref{app:QuarkSector}, and mesons, \Cref{app:MesonSector}. The flowing field $\dot\Phi$ and the related running of the Yukawa couplings is discussed in \Cref{app:QuarkMesonInterface}.

\subsubsection{Glue sector and quark-gluon interface}
\label{app:GlueSector}

The glue sector is treated as in \cite{Fu:2019hdw, Ihssen:2024miv, Fu:2024rto}. Still, for completeness, we briefly summarise here the calculation and approximation of the gluon propagator and the avatars of the strong coupling. The gluon propagator in the Landau gauge is parametrised as 
\begin{align}
	(G_A)^{ab}_{\mu\nu}(p) =\frac{1}{Z_A(p)\,p^2 \left[  1 +r_A(x) \right]}\Pi_{\mu \nu}^{\perp}\delta^{ab}\,,	
	\label{eq:GA} 
\end{align}
where the full momentum dependence is accounted for by the wave function $Z_A(p)$. For the explicit computation we we take 3d flat or Litim regulators with the shape function $r_A(x)= r_\textrm{flat}(x)$ with $x= \boldsymbol{p}^2/k^2$, see \labelcref{eq:Regs}. 

In perturbative QCD the running strong coupling can be extracted from different vertices and the RG-running of all these \textit{avatars} agree with each other up to two-loop due to the the Slavnov-Taylor identities (STIs) and two-loop universality for mass-independent RG-schemes. Beyond two-loop this does not hold even in perturbation theory and in a non-perturbative regime the STIs feature non-trivial scattering kernels that lead to non-trivial relations between the different avatars. Moreover, one has to consider all  tensor structures and the projections may not be orthogonal, for a more comprehensive discussion see e.g.~\cite{Dupuis:2020fhh, Pawlowski:2022oyq, Ihssen:2024miv}. Therefore one has to consider different avatars of the strong coupling that agree in the perturbative regime but deviate from each other in the infrared regime. We have the avatars 
\begin{align} 
	\alpha_{A^3}= \frac{g^2_{A^3}}{4 \pi}\,,\qquad \alpha_{A^4}= \frac{g^2_{A^4}}{4 \pi}\,,\qquad \alpha_{c\bar c A}= \frac{g^2_{c\bar c A}}{4 \pi}\,,\qquad \alpha_{l\bar l A}= \frac{g^2_{l \bar l A}}{4 \pi}\,,\qquad \qquad \alpha_{s\bar s A}= \frac{g^2_{s \bar s A}}{4 \pi}\,,  
\label{eq:AllAvatarsalpha}
\end{align}
where the subscripts indicate the scattering vertex, from which the avatar of $\alpha_s$ is extracted. In this work, we compute the $k$-dependence of the light and strange quark-gluon couplings and three-gluon coupling $\alpha_i=g_i^2 /(4 \pi)$ with $i=A^3, l\bar lA, s\bar s A$, and employ the approximation, 
\begin{align}
	\alpha_{A^4} \simeq  \alpha_{A^3}, \quad \alpha_{c \bar c A } \simeq \alpha_{l\bar l A}\,.
	\label{eq:alphaID}
\end{align}
The differences between all avatars $\alpha_i$ with $i=A^3,A^4,c\bar c A, l\bar lA, s\bar s A$ have been resolved in the vacuum, \cite{Mitter:2014wpa, Cyrol:2017ewj, Ihssen:2024miv, Fu:2025hcm} and at finite density, \cite{Pawlowski:2025jpg}. All avatars agree for momenta $p\gtrsim 2$\,GeV and the differences grow large for $p\lesssim 1$\,GeV. In the latter regime the glue dynamics decouples successively, which makes the approximation \labelcref{eq:alphaID} quantitatively reliable. 

We close this discussion of the strong couplings with a discussion of the full vertices and their dressings. All full vertices discussed above, $\Gamma^{(n_i})_i$ with $i=A^3, A^4, c\bar c A, q\bar q A$, admit more than one tensor structure. Their impact has been thoroughly investigated in pure Yang-Mills theory, \cite{Cyrol:2016tym, Cyrol:2017qkl} and in full QCD, \cite{Mitter:2014wpa, Cyrol:2017ewj, Fu:2025hcm}. Similar investigations have been performed in the DSE-method, see e.g.n~\cite{Gao:2021wun}.  In summary, for the present purpose the pure glue vertices, $i= A^3, A^4, c\bar c A$, are well approximated by their classical part. In turn, the transverse part of quark-gluon vertex is expanded in eight independent tensor structures, and three of them are important: the classical tensor structure and another chirally symmetric one, ${\cal T}_{A\bar{q}q}^{(1)}, {\cal T}_{A\bar{q}q}^{(7)}$ respectively, and a chirally breaking one, ${\cal T}_{A\bar{q}q}^{(4)}$. Depending on the chosen basis the two non-classical tensor structure take different forms. In the basis in \cite{Ihssen:2024miv} (Appendix F) we have 
\begin{align} 
\left[\mathcal{T}_{A\bar{q}q}^{(1)}\right]_\mu(p,q) 
=\,i\,\gamma_\mu\,,\qquad \left[\mathcal{T}_{A\bar{q}q}^{(4)}\right]_\mu(p,q) 
=\,i\,\sigma_{\mu\alpha}p_\alpha\,,\qquad \left[\mathcal{T}_{A\bar{q}q}^{(7)}\right]_\mu(p,q) 
=\frac{{\textrm{1}}}{3}\, \Big[ \sigma_{\alpha\beta}\gamma_\mu + \sigma_{\beta\mu}\gamma_\alpha + \sigma_{\mu\alpha}\gamma_\beta \Big]\,, 
\label{eq:QuarkGluonTensor}
\end{align}
with the dressings $\lambda^{(1,4,7)}_{q_i\bar q_i A}$. Within this set of three tensor structures the classical one is by far the most important, followed by the 4th tensor structure. Dropping the non-classical tensor structures, and in particular $\lambda^{(4,)}_{q_i\bar q_i A}$ in the DSE-approach can be compensated by a significant infrared enhancement of the dressing of the classical tensor structure, see e.g.~the recent review \cite{Fischer:2026uni}. In turn, in the fRG-approach with emergent composites, a significant part of this dynamics that is contained in the mesonic diagrams. Indeed, rewriting the loop hierarchy in the fRG in terms of DSEs, a significant part of the DSE of the vertex is carried by these loops and implicitly generate the vertex part with $\lambda^{(1)}_{q_i\bar q_i A}$. This entails that we can almost drop the tensor structures $\lambda^{(4,7)}_{q_i\bar q_i A}$ while still keeping quantitative precision. We are left with a small infrared enhancement factor of roughly $1\%$ which is details in \Cref{app:Pheno}. 

This setup is sufficient for the phase structure investigation in the present work. For the computation of thermodynamic observables such as pressure, entropy and the trace anomaly as well as fluctuations of conserved charges, a parametrisation with an additional mass function $m_A^2$ such as used in \cite{Pawlowski:2025jpg} is better suited and shows better convergence properties. While we will use such a parametrisation for the computation of the respective observables in \cite{Wen2026B-S}.  

The wave function $Z_A$ in 	\labelcref{eq:GA} is absorbed in the rescaled field $A_\mu$, and for the computation of the anomalous dimension $\eta_A$ in \labelcref{eq:etaPhi} we employ the same approximation as in \cite{Fu:2019hdw} which builds on \cite{Braun:2014ata}. There, the full anomalous dimension was expanded about that in the vacuum,  
\begin{align}
	\eta_{A}=\eta_{A,\textrm{vac}}^{\textrm{QCD}}+\Delta \eta_A^{\textrm{YM}} + \Delta \eta_A^{l}+\eta_A^{s}\,.
\end{align}
where $\eta_{A,\textrm{vac}}^{\textrm{QCD}}$ is an input from the $N_f=2$ flavour QCD precision results in the vacuum \cite{Cyrol:2017ewj}, and $\Delta \eta_A^{YM}$ is the finite-temperature contribution of Yang-Mills theory result \cite{Cyrol:2017qkl}. The term $\Delta \eta_A^{l}$ is the finite-temperature contribution of the light quark, and $\eta_A^{s}$ is the strange quark contribution. For the 3d flat or Litim regulator these contributions read with $i=l,s$,  
\begin{align}\nonumber 
	\eta_{A}^{i}
	&=-\frac{N_i}{\pi^2} g_{{\bar q}_i  A q_i}^{2}\int_0^1 d x
	\bigg[(1-\eta_{q_i}) \sqrt x+\eta_{q_i} x\bigg] \int_{-1}^1 d \cos \theta\bigg[
	\Big(\mathcal{FF}_{(1,1)}(m_{q_i},m_{q_i})-\mathcal{FF}_{(2,1)}(m_{q_i},m_{q_i})\Big) \\[1ex]
	&\hspace{4cm}
	+\Big(\sqrt x \cos^2\!\theta-\cos \theta\Big) \Big(1+r_F(x')\Big)\Big(\mathcal{FF}_{(2,1)}(m_{q_i},
	m_{q_i})-\frac{1}{2}\mathcal{FF}_{(1,1)}(m_{q_i},m_{q_i})
	\Big)\bigg]\,,
\end{align}
where the $r_{F}$ is the 3d flat shape function of the regulator of the quark. The multiplicities $N_i$ are given by $N_l=2$ and $N_s=1$. 

We proceed with the discussion of the flows of $\alpha_{l\bar l A},\alpha_{s\bar s A}$ and $\alpha_{A^3}$. For the light and strange quark-gluon couplings, we firstly consider their projection onto the classical tensor structures, 
\begin{align}
	\partial_t g_{q_i {\bar q}_i  A }=
	&\left( \frac12\eta_A+\eta_{q_i}\right) g_{q_i {\bar q}_i A }
	-\frac{1}{8(N_c^2-1)}\tr\,\left[\left[
	{{\textrm{Flow}}}^{(3)}_{ q_i\bar{q}_i A}\right]^{a}_{\mu}
	\left[ T^{{(1)}}_{ q_i{\bar q}_i  A}\right]^a_{\mu}\right](\{p\})\,,
\label{eq:dtg_q}
\end{align}
The diagrammatic part of the flow of the quark-gluon couplings on the right hand side of \labelcref{eq:dtg_q} receives contributions from quark-gluon diagrams and quark-meson diagrams,
\begin{align} \partial_t g_{q_i {\bar q}_i A }
	= \left( \frac12\eta_A+\eta_{q_i}\right)
	g_{q_i {\bar q}_i A }+\bigg({\textrm{Flow}}^{(3),A}_{(q_i {\bar q}_i  A
		)} + {\textrm{Flow}}^{(3),\phi}_{{(q_i {\bar q}_i  A
			)}}\bigg)\,, 
\end{align}
where the parenthesis in the subscript indicates the projection procedure defined in 
\labelcref{eq:dtg_q}. The quark-gluon diagrams have the form 
\begin{align}
	{\textrm{Flow}}^{(3),A}_{({q_i {\bar q}_i A })}
	=&\frac{3}{8\pi^2N_c}
	g_{q_i {\bar q}_i A }^{3}\frac{{m}_{q_i}^{2}}{k^2}\bigg
	\{\frac{2}{15}(5-\eta_{A})\mathcal{FB}_{(2,2)}(
	m_{q_i},0)+\frac{1}{3}(4-\eta_{q_i})\mathcal{FB}_{(3,1)}(
	m_{q_i},0)\bigg\}\nonumber \\[1ex]
	&+\frac{3N_c}{8\pi^2}g_{q_i {\bar q}_i A }^{2}g_{A^3}
	\bigg\{\frac{1}{20} (5-\eta_{q_i})
	\mathcal{FB}_{(1,2)}(m_{q_i},0) -\frac{1}{6} (4-\eta_{q_i})\mathcal{FB}_{(2,1)}(m_{q_i},0)\nonumber \\[1ex]
	&  +\frac{1}{30} (5-2\eta_{q_i}) \mathcal{FB}_{(2,2)}(m_{q_i},0)
	-\frac{4}{15}(5-\eta_{A})\mathcal{FB}_{(1,2)}(m_{q_i},0)\nonumber \\[1ex]
	&  +\frac{1}{30}(10-3\eta_{A}) \mathcal{FB}_{(1,3)}(m_{q_i},0)\bigg\} \,.
\end{align}
The light quark-meson diagrams have the form 
\begin{align}
	{\textrm{Flow}}^{(3),\phi}_{{({l \bar{l}A })}}
	=& -\frac{1}{2\pi^2}\bigg(\frac{g_{l \bar{l}A } {h}_{l}^{2}}{8}\bigg(3\mathcal{L}_2(\pi,q_l)+3\mathcal{L}_2(a_0,q_l)   
	+\cos^2 \varphi_s\mathcal{L}_2(f_0,q_l) +\sin^2 \varphi_s\mathcal{L}_2(\sigma,q_l)
	\nonumber \\[1ex]
	&+\cos^2 \varphi_p \mathcal{L}_2(\eta,q_l)+\sin^2 \varphi_p \mathcal{L}_2(\eta ',q_l)\bigg)
	+\frac{ g_{s \bar{s}A } {h}_{l}{h}_{s}}{4}\bigg(\mathcal{L}_2(\kappa,q_s)+\mathcal{L}_2(K,q_s)\bigg)\bigg)
\end{align}
where we introduced 
\begin{align}\nonumber
	\mathcal{L}_2(\phi_i,q_i)&=\frac{2}{3}\Biggl\{\bigg(1-\frac{\eta_{q_i}}{4}\bigg)\left[\mathcal{FB}_{2,1}(m_{q_i},m_{\phi_i})+2\frac{m_{q_i}^2}{k^2}\mathcal{FB}_{3,1}(m_{q_i},m_{\phi_i})\right] \\[1ex] 
	&+\bigg(1-\frac{\eta_\phi}{5}\left[\bigg(\mathcal{FB}_{(1,2)}(m_{q_i},m_{\phi_i})+\frac{m_{q_i}^2}{k^2}\mathcal{FB}_{(2,2)}(m_{q_i},m_{\phi_i})\right]\Biggr\}\,.
\end{align}
The strange quark-gluon diagrams have the form 
\begin{align}
	{\textrm{Flow}}^{(3),\phi}_{{({s \bar{s}A })}}
	=& -\frac{1}{2\pi^2}\bigg(\frac{g_{l \bar{l}A } {h}_{l}{h}_{s}}{4}\bigg(\mathcal{L}_2(\kappa,q_s)+\mathcal{L}_2(K,q_s)\bigg)
	+  \frac{g_{s \bar{s}A } {h}_{s}^2}{4}(\sin^2 \varphi_s \mathcal{L}_2(f_0,q_s)+\cos^2 \varphi_s\mathcal{L}_2(\sigma,q_s)
	\nonumber \\[1ex]  
	&+\sin^2 \varphi_p\mathcal{L}_2(\eta,q_s)+\cos^2 \varphi_p  \mathcal{L}_2(\eta ',q_s)
	\bigg)\bigg)\,.
\end{align}
We have already discussed below \labelcref{eq:QuarkGluonTensor}, that the quark-gluon couplings are augmented with a small infrared enhancement of roughly $1\%$, which is discussed in \Cref{eq:IR-Enhancement}.  

Finally, we discuss the flow of $\alpha_{A^^3}$. Its flow is expanded about the vacuum result similarly to the gluon anomalous dimension, 
\begin{align}
	\partial_t g_{A^3}=\partial_t g_{A^3, \textrm{vac}}+\partial_t \Delta g_{A^3}\,,
\end{align}
For the vacuum contribution, we use \cite{Braun:2014ata} as input, while the finite-temperature contribution is approximated as
\begin{align}
	\partial_t \Delta g_{A^3} = \partial_t \Delta g_{l \bar l A}\,.
\end{align}
This concludes our discussion of the glue sector.

\subsubsection{Meson sector}
\label{app:MesonSector}

The flow of the chiral part of the effective potential reads
\begin{align}\label{eq:flowU}
\partial_t \tilde{V}_{\Sigma,k} (\Sigma,\Sigma^\dagger)&= \frac{k^4}{4 \pi^2}\bigg\{ \sum_i l_0^{(B,4)}(m_{\phi_i},\eta_{\phi};T,\mu_i)
-4 N_c\big[ 2 l_0^{(F,4)}(m_{l},\eta_l;T,\mu)
 + l_0^{(F,4)}(m_{s},\eta_s;T,\mu_s)\big]\bigg\},
\end{align}
where the threshold functions $l_0^{(B/F,4)}$ are given in \cite{Fu:2019hdw}. The summation runs over all of the scalar and pseudoscalar nonet meson fields.

The mesonic wave function $\boldsymbol{Z}_\phi$ is a diagonal matrix with entries $Z_{\phi_i}$. In QCD, these wave functions run over orders of magnitude in the perturbative regime while they change little in the infrared below the symmetry breaking scale $k_\chi\approx 400$\,MeV, see \cite{Ihssen:2024miv}. The latter finding in QCD is corroborated in low energy effective theories, for a comprehensive compilation of works see \cite{Dupuis:2020fhh}. This suggests to identify them and we only compute the pion wave function as it is the softest mesonic mode for all $T$ and $\mu_B$ except a very small regime around the CEP, 
\begin{align} 
	Z_{\phi_i}=Z_\phi:=Z_\pi\,,\quad \forall i \quad \longrightarrow \eta_{\phi_i}=\eta_\phi:=\eta_\pi\,.
\label{eq:OneZphi}
\end{align}
Then, the meson-loop contributions to the uniform meson anomalous dimension are given as
\begin{align}
\eta^{\textrm{meson}}_{\phi}&=\frac{1}{3 \pi^2} \bigg [  V_{\pi^{+} \pi^{-} f_0} \mathcal{BB}_{(2,2)} (m_{\pi},m_{f_0})+  V_{\pi^{+} \pi^{-} \sigma}\mathcal{BB}_{(2,2)} (m_{\pi},m_{\sigma})  +  V_{\pi^{+} a_0^{-} \eta}\mathcal{BB}_{(2,2)} (m_{a_0},m_{\eta})\nonumber\\[1ex]
&\quad+ V_{\pi^{+} a_0^{-} \eta'}\mathcal{BB}_{(2,2)} (m_{a_0},m_{\eta'}) 
+(V_{\pi^{+} \kappa^{-} K^0}+V_{\pi^{+}K^{-}\kappa^0}) \mathcal{BB}_{(2,2)} (m_{\kappa},m_{K}) \bigg ]\,,\nonumber
\end{align}
where the expressions of the threshold functions $\mathcal{BB}_{(m,n)}$ as well as those used in the following, e.g., $\mathcal{F}_{(n)}$, $\mathcal{FF}_{(m,n)}$, $\mathcal{FB}_{(m,n)}$, can be found in \cite{Fu:2019hdw}. The quark-loop contribution to the meson anomalous dimension at $p=0$ reads
\begin{align}
\eta^\textrm{quark}_{\phi}(0,0)= \frac{N_{c}{h}_{l}^{2}}{6\pi^2}\bigg[(2\eta_{l}-3)\mathcal{F}_{(2)}(
    {m}_{l}) -4(\eta_{l}-2)\mathcal{F}_{(3)}({m}_{l})\bigg]\,.
\end{align}
The quark-loop contribution to the meson anomalous dimension at $\bm p^2= k^2$ and $p_0^2=0$ is given by
\begin{align}
\eta^\textrm{quark}_{\phi}(0,\bm k)&= -\frac{N_{c}}{\pi^2} h_l^2\int_0^1 d x\bigg[(1-\eta_{l}) \sqrt x
    +\eta_{l} x\bigg]\nonumber\\
  &\quad\times \int_{-1}^1 d \cos \theta\Bigg\{\bigg[
    \Big(\mathcal{FF}_{(1,1)}({m}_{l},
   {m}_{l})-\mathcal{F}_{(2)}(
    {m}_{l}) \Big)-\Big(\mathcal{FF}_{(2,1)}({m}_{l},
    {m}_{l})-\mathcal{F}_{(3)}({m}_{l})\Big)\bigg] \nonumber \\[1ex]
  &\quad+\bigg[\Big(\sqrt x -\cos \theta\Big)
    \Big(1+r_F(x')\Big)\mathcal{FF}_{(2,1)}(
    {m}_{l},{m}_{l})-\mathcal{F}_{(3)}({m}_{l})\bigg]\nonumber \\[1ex]
  &\quad -\frac{1}{2}\bigg[\Big(\sqrt x -\cos \theta\Big)
    \Big(1+r_F(x')\Big)\mathcal{FF}_{(1,1)}({m}_{l},{m}_{l}) -\mathcal{F}_{(2)}({m}_{l})\bigg]\Bigg\}\,.
\end{align}
This concludes our discussion of the meson sector.

\subsubsection{Quark Sector}
\label{app:QuarkSector}

In this work, we distinguish the light and strange quark wave functions, i.e., $Z_q = \mathrm{diag}(Z_l, Z_l, Z_s)$. In principle, each of the mesons has its respective wave function. However, for the sake of simplicity, we do not distinguish them and assume $Z_\Sigma=Z_{\pi^+}$, which is also employed in the 2+1 flavour quark-meson model~\cite{Rennecke:2016tkm}. 

We distinguish the light and strange quark anomalous dimensions, which can be calculate from the two-point quark correlation functions with the lowest momentum and frequency \cite{Pawlowski:2014zaa, Fu:2015naa, Rennecke:2016tkm, Fu:2019hdw}, viz.,
\begin{align}
\eta_{q_i}=\frac{1}{4 Z_{q_i}} \text{Re} \Bigg [ i \frac{\partial}{\partial |\bm{p}|^2} \text{tr}\left( \bm \gamma \cdot \bm p\, \partial_t \Gamma_{{\bar q}_i  q_i}^{(2)} (p) \right)\Bigg ]_{\bm p=0},
\end{align}
without summation over the flavour index $i$. Then one finds for the light quark anomalous dimension
\begin{align}
\eta_{l}&=\frac{1}{48 \pi^2} (4-\eta_{\phi})h_{l}^2 \Big [ 3 \mathcal {FB}_{(1,2)}(m_l,m_\pi)+3 \mathcal {FB}_{(1,2)}(m_l,m_{a_0})+\cos^2\varphi_s \mathcal {FB}_{(1,2)}(m_l,m_{f0})\nonumber\\[1ex]
&\quad+\cos^2\varphi_p \mathcal {FB}_{(1,2)}(m_l,m_{\eta})+\sin^2\varphi_s \mathcal {FB}_{(1,2)}(m_l,m_{\sigma})+\sin^2\varphi_p \mathcal {FB}_{(1,2)}(m_l,m_{\eta'}) \Big]\nonumber\\[1ex]
&\quad+\frac{1}{24 \pi^2} \frac{N_c^2-1}{2 N_c} g_{l\bar l A}^2 \Big [ 2(4-\eta_{A})\mathcal {FB}_{(1,2)}(m_l,0)+3(3-\eta_{l})\big(\mathcal{FB}_{(1,1)}(m_l,0)-2 \mathcal{FB}_{(2,1)}(m_l,0)\big )\Big ],
\end{align}
and for the strange quark anomalous dimension
\begin{align}
\eta_{s}&=\frac{1}{24 \pi^2} (4-\eta_{\phi})h_{s}^2 \Big [\sin^2\varphi_s \mathcal {FB}_{(1,2)}(m_s,m_{f_0}) +\sin^2\varphi_p \mathcal {FB}_{(1,2)}(m_s,m_{\eta})+\cos^2\varphi_s \mathcal {FB}_{(1,2)}(m_s,m_{\sigma}) \nonumber\\[1ex]
&\quad+\cos^2\varphi_p \mathcal {FB}_{(1,2)}(m_s,m_{\eta'}) \Big ]\nonumber\\[1ex]
&\quad+\frac{1}{24 \pi^2} \frac{N_c^2-1}{2 N_c} g_{s\bar s A}^2 \Big [ 2(4-\eta_{A})\mathcal {FB}_{(1,2)}(m_s,0)+3(3-\eta_{s})\big(\mathcal{FB}_{(1,1)}(m_s,0)-2 \mathcal{FB}_{(2,1)}(m_s,0)\big) \Big ].
\end{align}
This concludes our discussion of the quark sector.

\subsubsection{Quark-meson interface and emergent composites}
\label{app:QuarkMesonInterface}

In this Supplement we discuss the absorption of the four-quark scattering in the scalar-pseudoscalar sector into the scattering with emergent composites. This also allows us to take into account multi-scattering events of resonant soft modes such as the pion. The flowing field $\dot\Phi$ only has components in the light and strange sector, 
\begin{align}
\dot \phi[\Phi]&=(\bar q \, \dot{\boldsymbol{A}}^{\frac{1}{2}} T^a  \dot{\boldsymbol{A}}^{\frac{1}{2}} \,  q\,, \,\bar q\, \dot{\boldsymbol{A}}^{\frac{1}{2}} i \gamma _5 T^a \dot{\boldsymbol{A}}^{\frac{1}{2}}\, q\,)\,,
\label{eq:dotphi} 
\end{align}
with the diagonal matrix $\dot{\boldsymbol{A}}_{k}$, also called the hadronisation function, 
\begin{align}
\dot{\boldsymbol{A}}_{k}=\begin{pmatrix}
\dot{A}_{l,k}&0&0\\
0&\dot{A}_{l,k}&0\\
0&0&\dot{A }_{s,k}
\end{pmatrix}\,.
\end{align}
For convenience, we introduce light and strange matrices $T^l$ and $T^s$, 
\begin{align}
T^l=\frac{1}{2}
\begin{pmatrix}
1& 0 & 0\\
0 & 1 & 0 \\
0 & 0 & 0
\end{pmatrix}\,,
\qquad
T^s=\frac{1}{\sqrt{2}}
\begin{pmatrix}
0& 0 & 0\\
0 & 0 & 0 \\
0 & 0 & 1
\end{pmatrix}\,.
\end{align}
Now we take a four-quark derivative of the generalised flow equation \labelcref{eq:GenFlow} and project onto the 
scalar-pseudoscalar sector. Heuristically this can be understood as the second  $(\bar q\, T^i q)$-derivative of \labelcref{eq:GenFlow} with $i=s,l$. This leads us to 
\begin{align}
- \partial_t \lambda_{i} + 2\eta_{q _{i}}\lambda_{i} + \dot{A}_{i,k} h_{i,k}=-\text{Flow} _{(\bar q T^i q) (\bar q T^i q)}^{(4)},
\qquad \text{with} \qquad i=l,s\,.
\end{align}
Here, $\text{Flow} _{(\bar q T^i q) (\bar q T^i q)}^{(4)}$ denotes the projection of the four-quark flow onto the respective tensor structure with the coupling  $\lambda_i$. We determine $\dot{\boldsymbol{A}}$ such that the four-quark couplings $\lambda_{l,s}$ vanish, 
\begin{align}
\lambda_{i}  \equiv 0\, , \quad \forall k\,, \qquad i=l,s\,.
\label{lambda_eq}
\end{align}
\Cref{lambda_eq} is obtained with
\begin{align}\nonumber 
\dot{A}_{l,k}=-\frac{1}{ h_{l}} \text{Flow}_{(\bar q T^l q)^2}^{(4)}\,, \\[1ex]
\dot{A}_{s,k}=-\frac{1}{ h_{s}}  \text{Flow}_{(\bar q T^s q)^2}^{(4)}\,.
\end{align}
For the Yukawa couplings $h_i$ we use that they can be obtained from the mass functions by dividing out $\sigma_i$ with $i=l,s$. Hence we project the bilinear quark derivative of \labelcref{eq:GenFlow} onto its scalar part at vanishing momentum and multiply it with $1/\sigma_l$ and $\sqrt{2}/\sigma_s$ respectively. Heuristically these are derivatives with respect to $(\bar q\, T^i q)$, multiplied with $1/\sigma_i$. With this procedure we arrive at 
\begin{align}
	\partial_t h_{i}&=\bigg(\eta_{i}+\frac{1}{2} \eta_{\phi_i}\bigg) h_{i} - \frac{1}{ \sigma_i}\frac{\partial \left( {V}_\phi(\phi)+ c_i\, \sigma_i\right)}{\partial  \sigma_i} \dot{ A}_{i}+\frac{1}{ \sigma_i}\text{Re}\left[({\text{Flow}}^{(2)}_{(\bar q T^i q)}\right]\,.  \label{eq:Yukawa_R}
\end{align}
In the present work we use $\eta_{\phi_i}=\eta_\phi$, see \labelcref{eq:etaPhi} and the respective discussion. 

The combination ${V}_\phi(\phi)+ c_i\, \sigma_i$ removes the explicit breaking term from the flow, see \labelcref{eq:GenFlow} and leads to manifest chiral invariance of the flow in the absence of the anomalous breaking of chiral invariance. 
This term has the explicit form 
\begin{align}\nonumber 
	\frac{1}{ \sigma_l}\frac{\partial \left( {V}_\phi(\phi)+ c_l\, \sigma_l\right)}{\partial  \sigma_l}&=- \frac{ c_A  \sigma_s}{\sqrt{2}}  + V_\chi^{(1,0)} - V_\chi^{(0,1)}\,,\\[1ex]
\frac{1}{ \sigma_s}\frac{\partial \left( {V}_\phi(\phi)+ c_s\, \sigma_s\right)}{\partial  \sigma_s}&=- \frac{ c_A  \sigma_l^2}{2 \sqrt{2}  \sigma_s}+ V_\chi^{(1,0)} +2  V_\chi^{(0,1)}\,.
\label{eq:CovariantShifts}
\end{align}
Note that $({1}/{ \sigma_l}) \,{\partial \left( {V}_\phi(\phi)+ c_l\, \sigma_l\right)}/{\partial  \sigma_l}$ is nothing but the pion mass $ m_\pi^2$ as in the two flavour case. 

The last term on the right hand side of the light flavour case in the first line of  \labelcref{eq:Yukawa_R} reads
\begin{align}
\frac{1}{\bar \sigma_l}\text{Re} ({\text{Flow}}_{(\bar q T^l q)}^{(2)})&=
\frac{1}{8 \pi^2} h_l^3\big [ 3 \mathcal{L}_{(1,1)}(m_l,m_{a_0})-3 \mathcal{L}_{(1,1)}(m_l,m_{\pi})+\cos^2 \varphi_s \mathcal{L}_{(1,1)}(m_l,m_{f_0})\nonumber\\[1ex]
&\quad-\cos^2 \varphi_p \mathcal{L}_{(1,1)}(m_l,m_{\eta})+\sin^2 \varphi_s \mathcal{L}_{(1,1)}(m_l,m_{\sigma})-\sin^2 \varphi_p \mathcal{L}_{(1,1)}(m_l,m_{\eta'})]\nonumber\\[1ex]
&\quad-\frac{3}{2 \pi^2} \frac{N_c^2-1}{2 N_c} g^2 _{l\bar l A}  h_l \mathcal{L}_{(1,1)}(m_l,0)\,.
\end{align}
The last term on the right hand side of the strange flavour case in the second line of  \labelcref{eq:Yukawa_R} reads
\begin{align}
\frac{1}{ \sigma_s}\text{Re} ( {\text{Flow}}_{(\bar q T^s q)}^{(2)})&=
\frac{1}{4 \pi^2} h_s^3\big [ \sin^2 \varphi_s \mathcal{L}_{(1,1)}(m_s,m_{f_0}) -\sin^2 \varphi_p \mathcal{L}_{(1,1)}(m_s,m_{\eta})+\cos^2 \varphi_s \mathcal{L}_{(1,1)}(m_s,m_{\sigma})\nonumber\\[1ex]
&\quad -\cos^2 \varphi_p \mathcal{L}_{(1,1)}(m_s,m_{\eta'})]-\frac{3}{2 \pi^2} \frac{N_c^2-1}{2 N_c} g^2 _{s\bar s A}  h_s \mathcal{L}_{(1,1)}(m_s,0)\,.
\end{align}
In our calculations, we distinguish the four-quark flows for the light and strange quarks. The gluon exchange diagrams are given by
\begin{align}
{\text{Flow}}^{(4),A}_{(\bar q T^{i} q)^2}=&-\frac{3}{2\pi^2}\frac{1}{k^2}\frac{N_c^2-1}{2 N_c}\left(\frac{3}{4} -\frac{1}{N_c^2}\right)g_{q_{i} \bar q_{i} A}^4 \bigg \{ \frac{2}{15} (5-\eta_{A}) [\mathcal{FB}_{(1,3)}(m_{q_i},0)\nonumber\\[1ex]
&- (m_{q_i}/k)^2 \mathcal{FB}_{(2,3)}(m_{q_i},0)]+\frac{1}{12}(4-\eta_{q_i})[\mathcal{FB}_{(2,2)}(m_{q_i},0)]-2 ( m_{q_i}/k)^2 \mathcal{FB}_{(3,2)}(m_{q_i},0)]\bigg\}.
\end{align}
In the computation of  the meson exchange diagrams, we have ignore the mixing angles between the light-strange basis and the physical basis. The expression of meson exchange diagrams for light quark are the same as that in \cite{Fu:2019hdw}.  For the strange quark, the meson exchange diagrams read
\begin{align}
{\text{Flow}}^{(4),\phi}_{(\bar q T^{s} q)^2}
&=\frac{1}{8\pi^2}\frac{1}{k^2}\frac{1}{N_c}
    {h}_{s}^{4}\bigg\{\frac{2}{15}(5-\eta_{\phi})\Big[\Big(\mathcal{FBB}_{(1,1,2)}(
    {m}_{s}^{2},{m}_{\eta_s}^{2},{m}_{\sigma_s}^{2})+\mathcal{FBB}_{(1,1,2)}({m}_{s}^{2},
    {m}_{\sigma_s}^{2},{m}_{\eta_s}^{2})\nonumber \\[1ex] 
&\quad-2\mathcal{FB}_{(1,3)}({m}_{s}^{2},
    {m}_{\eta_s}^{2})\Big)-({m}_{s}/k)^{2}\Big(
    \mathcal{FBB}_{(2,1,2)}({m}_{s}^{2},
    {m}_{\eta_s}^{2},m_{\sigma_s}^{2})
+\mathcal{FBB}_{(2,2,1)}({m}_{s}^{2},{m}_{\eta_s}^{2},
    {m}_{\sigma_s}^{2})\nonumber \\[1ex] 
&\quad-2\mathcal{FB}_{(2,3)}({m}_{s}^{2},{m}_{\eta_s}^{2})\Big)\Big]
    +\frac{1}{6}(4-\eta_{s})\Big[\Big(\mathcal{FBB}_{(2,1,1)}({m}_{s}^{2},
    {m}_{\eta_s}^{2},{m}_{\sigma_s}^{2})-\mathcal{FB}_{(2,2)}({m}_{s}^{2},{m}_{\eta_s}^{2})
    \Big)\nonumber \\[1ex] 
&\quad-2({m}_{s}/k)^{2}\Big(\mathcal{FBB}_{(3,1,1)}({m}_{s}^{2},
    {m}_{\eta_s}^{2},{m}_{\sigma_s}^{2})-\mathcal{FB}_{(3,2)}({m}_{s}^{2},{m}_{\eta_s}^{2})\Big)\Big]\bigg\}\,.
\end{align}
This concludes our discussion of the quark-meson interface.

\subsection{Physics Parameters at the initial cutoff scale} 
\label{app:set-up}

In functional QCD, the only parameters to fix are that of fundamental QCD. In the 2+1 flavour QCD case discussed here, these are only the current quark masses $m_i^0$ with $i=l,s$ at the initial infrared cutoff scale chosen deep in the UV, $k=\Lambda_\textrm{UV}$. The current quark masses have to be adjusted such that physical QCD is approached in the infrared. This simply means that $m_i^0, m_s^0$ are fixed by two appropriately chosen two observables, such as the ratios $m_\pi/f_\pi, m_K/f_\pi$. For a comprehensive discussion see in particular \cite{Ihssen:2024miv, Fu:2025hcm}. Such a procedure amounts to the following: we use a small strong coupling at the initial scale $\Lambda_\textrm{UV}$. Then the effective action at $k=0$ is computed with the flow equation. The correct ratios $m_\pi/f_\pi, m_K/f_\pi$ are obtained by fine-tuning the initial conditions and rerunning the flow until convergence. This procedure has been done in \cite{Ihssen:2024miv, Fu:2025hcm} and we refer to these works for more details. Furthermore, it has been checked in \cite{Fu:2025hcm} that an initial scale  
$\Lambda_\textrm{UV}\approx 20$\,GeV allows for using a classical approximation of the effective action at this scale: all non-trivial parts of the initial effective action are set to zero at this scale, leaving us only with the parameters of fundamental QCD.

\subsubsection{QCD parameters at the initial cutoff scale} 
\label{app:QCDparameters}

With the knowledge and results from  \cite{Fu:2019hdw, Ihssen:2024miv, Fu:2025hcm} we avoid this fine-tuning and resort to the following procedure. We use $\Lambda=20$\,GeV in the physical units there and use the respective initial strong coupling from \cite{Fu:2019hdw} for all avatars, 
\begin{align}
\alpha_{s,\Lambda_\textrm{UV}}=0.235\,, \qquad \Lambda=20\,\textrm{GeV}\,. 
\label{eq:alphaInitial}
\end{align}
The validity of this initial condition is then checked with the results for the meson masses, see \Cref{tab:Parameters+Observables}. We use a Gau\ss ian initial effective potential 
\begin{align}
 V_{\phi,\Lambda_\textrm{UV}}(\rho_1,\rho_2)= m_{\Lambda,\phi}^2  \rho_1 -c_A\,\xi-  c_l\,\sigma_l- c_s\,\sigma_s\,,
 \label{eq:InitialVphi}
\end{align}
with $m_{\Lambda,\phi}^2=10^2\times\Lambda^2$ $\mathrm{GeV}^{2}$, and the initial Yukawa couplings are given by 
\begin{align} 
	h_{i, \Lambda_\textrm{UV}}= 1\,,\qquad i=l,s\,.
\label{eq:InitialYukawas}
\end{align} 
With \labelcref{eq:InitialVphi,eq:InitialYukawas}, the respective four-quark couplings are given by 
\begin{align} 
\lambda_i \Lambda_\textrm{UV}^2= \frac12 \frac{h_{i,\Lambda_\textrm{UV}}^2}{ m_{\phi,\Lambda_\textrm{UV}}}  = 0.05 ,,\qquad i=l,s\,.
\label{eq:Initial4Quark}
\end{align} 
\Cref{eq:Initial4Quark} entails that the four-quark coupling is close vanishing, avoiding the gauge-NJL regime of the present approach. It has been checked in the literature, that large variations of these initial conditions do not change the results at $k=0$, if the initial four-quark coupling is negligible: The value of the four-quark coupling at $k=0$ is completely dominated by the flow.     

\begin{table}[t]
	\begin{center}
		\begin{tabular}{|c|c||c|c|}
			\hline & & & \\[-2ex]
			Vacuum Observables  & Value & Parameters & Value  \\[1ex]
			\hline & & &   \\[-2ex]
			$m_{\pi}$ & \, 140 MeV \, & $c_{l}$ & \, $4.10\times 10^{8}\,\mathrm{MeV}^{3}$ \, \\[1ex]
			\hline & & & \\[-2ex] 
			$m_{K}$ & 480 MeV & $c_{s}$ & $1.12 \times 10^{10}\,\mathrm{MeV}^{3}$ \\[1ex]
			\hline & & & \\[-2ex] 
			$m_{\eta}$ & 526 MeV & $ c_{A0}$ & $2.6\times 10^{4}$ $\mathrm{MeV}$ \\[1ex]
			\hline & & & \\[-2ex] 
			$m_{\eta'}$ & 965 MeV & $ k_{0}$ & 700 MeV \\[1ex]
			\hline & & & \\[-2ex] 
			$M_{l}$ & 347 MeV &  &  \\[1ex]
			\hline & & & \\[-2ex] 
			$M_{s}$ & 464 MeV &  &  \\[1ex]
			\hline
		\end{tabular}
		\caption{The parameters of explicit chiral symmetry breaking, the parameters of ’t Hooft coupling, and the vacuum observables. The strong coupling at the initial cutoff scale is chosen as $\alpha_{s,\Lambda}=0.235$, see \labelcref{eq:alphaInitial}. This implies an initial cutoff scale of $\Lambda=20$\,GeV.}
		\label{tab:Parameters+Observables}
	\end{center}\vspace{-0.5cm}
\end{table}

The current quark masses $m_l^0,m_s^0$ should be determined by the parameters of explicit chiral symmetry breaking, i.e., $c_l$ and $c_s$. In the present setup this ratio is identical to the ratio of current quark masses: $c_s/c_l= m_{s}^0/m_{l}^0$, see \cite{Fu:2019hdw}. As already discussed above, $m_l^0,m_s^0$  and hence $c_l,c_s$ have to be chosen such that they reproduce $m_\pi,m_K$ for $k=0$. In the present work we avoid the fine-tuning by using the information from \cite{Fu:2019hdw}: the ratio of $c_s/c_l = 27.40$, and we simply tune $c_l$ such that the pion mass is given by $m_\pi=140$\,MeV. 

All parameters and the respective observables are listed in \Cref{tab:Parameters+Observables}. We note in passing that the ratio of the current quark masses is given by the ratio $c_s/c_l$. Hence we have $m_{s}^0/m_{l}^0=27.40$, which is very close to the ratio of current quark masses on the lattice:   $m_{s,\textrm{lat}}^0/m_{l,\textrm{lat}}^0=27.42(12)$ in \cite{FlavourLatticeAveragingGroupFLAG:2024oxs}.

\subsubsection{Phenomenological parameters} 
\label{app:Pheno}

In this work, we do not resolve the full QCD dynamics. Specifically, we have to compensate for dropped tensor structures in the quark-gluon vertex, the only qualitative resolution of the running of the 't Hooft determinant computed in \cite{Pawlowski:1996ch}, and the approximate resolution of the Polyakov loop potential within functional QCD, see \cite{Haas:2013qwp}. \\[-2ex]

\paragraph{Infrared enhancement of the quark-gluon coupling} 
\label{app:IR-enhancement}

We compensate for the dropped tensor structures of the quark-gluon vertex with a small infrared enhancement of the quark-gluon coupling of roughly 1\%, see \labelcref{eq:IR-Enhancement}. 

To compensate for the missing non-classical tensor structures of the quark-gluon vertices, we use an infrared enhancement as that used in \cite{Braun:2014ata,Fu:2019hdw, Ihssen:2024miv},  
\begin{align}
	\varsigma_{a,b}(k) = 1+ a\, \frac{(k/b)^\delta}{e^{(k/b)^\delta}-1}\,, 
	\label{eq:IR-Enhancement}
\end{align}
Here, $b=2$\,GeV and $\delta=2$, chosen to be the same as in \cite{Fu:2019hdw}. The enhancement strength parameter $a=0.009$, which is adjusted to obtain the vacuum constituent light quark mass $m_{l,k=0}=347$ MeV.\\[-2ex]

\paragraph{'t Hooft determinant} 
\label{app:thooft}

we introduce an instanton-induced 't Hooft coupling for $k \gtrsim 700$ MeV \cite{Pawlowski:1996ch} and $N_f=3$,
\begin{align}
 c_{A,k} =c_{A0} \left[ e^{- \left( \frac{2 \pi}{\alpha_{s,k}}-  \frac{2 \pi}{\alpha_{s,k_0} }\right) }\, \left(\frac{k_0}{k}\right)^{5}\,\theta(k-k_0) +
  \frac{1 }{e^{\frac{k-k_{0}}{\Delta_k}}+1}\,\theta(k_0-k)\right]\,, \qquad \textrm{with}\qquad  \alpha_{s,k}=\alpha_{l\bar lA,k}(T=0)\,,
 \label{eq:cAdefinition}
 \end{align}
which has the correct high energy behaviour computed in \cite{Pawlowski:1996ch}. In the the low-energy regime the $U_A(1)$ breaking parameter $c_{A,k}$ saturates. The free parameters $k_{0}=700$ MeV, $c_{A0}=2.6\times 10^4$ MeV are determined by the masses of $\eta$ and $\eta'$ meson in the vacuum.\\[-2ex]

\paragraph{Polyakov loop potential} 
\label{app:PolPot}

We use the approximation to the full Polyakov loop potential suggested in \cite{Haas:2013qwp}. In this approach the parameters of a Polyakov potential obtained from lattice input are determined from functional QCD.  

For the lattice input we take the parametrisation of the Polyakov-loop potential from \cite{Lo:2013hla}, which includes the longitudinal and transverse fluctuations of the Polyakov loop. A comparison of the equation of state between different parametrisations of the Polyakov-loop potential within low-energy effective theories is given in \cite{Wen:2018nkn}. The parametrisation of the Polyakov-loop potential reads
\begin{align}
   \nonumber V_{A_0}(A_0) &= -\frac{\bar a(T)}{2} \bar L L + \bar b(T)\ln M_H(L,\bar{L})\\ 
  &\quad+ \frac{\bar c(T)}{2} (L^3+\bar L^3) + \bar d(T) (\bar{L} L)^2,
\end{align}
with $L=L(A_0)$ and 
\begin{align}
M_H (L, \bar{L})&= 1 -6 \bar{L}L + 4 (L^3+\bar{L}^3) - 3  (\bar{L}L)^2\,.
\end{align}
for $x\in \{\bar a, \bar c, \bar d\}$
\begin{align}\label{eq:polyakov_xT}
  x(T) &= \frac{x_1 + x_2/(t+1) + x_3/(t+1)^2}{1 + x_4/(t+1) + x_5/(t+1)^2}\,,
\end{align}
and
\begin{align}\label{eq:polyakov_bT}
  \bar b(T) &=\bar b_1 (t+1)^{-\bar b_4}(1 -e^{\bar b_2/(t+1)^{\bar b_3}} )\,.
\end{align}
The relevant coefficients are listed in \Cref{tab:polyakov}. The reduced temperature of QCD is given by
\begin{align}
  t_{\text{YM}}\rightarrow \alpha\,t_{\text{glue}},\quad
  t_{\text{glue}}\equiv \frac{T-T_c^\text{glue}}{T_c^\text{glue}}\,,
\end{align}
where we choose
\begin{align}
T_c^{\text{glue}}=285\, \text{MeV},\quad
\alpha=0.60\,.
\end{align}
This concludes the discussion of the phenomenological parameters.

\begin{table}[t]
  \centering
  \begin{tabular}{|c||c|c|c|c|c|}
    \hline
    & 1 & 2 & 3 & 4 & 5 \rule{0pt}{2.6ex}\rule[-1.2ex]{0pt}{0pt}
    \\[1ex] \hline & & & & & \\[-1ex]
    $\bar a_i$ &-44.14& 151.4 & -90.0677 &2.77173 &3.56403 \\[1ex]  & & & & & \\[-2ex]
    $\bar b_i$ &-0.32665 &-82.9823 &3.0 &5.85559  & \\[1ex]  & & & & & \\[-2ex]
    $\bar c_i$ &-50.7961 &114.038 &-89.4596 &3.08718 &6.72812 \\[1ex]  & & & & & \\[-2ex]
    $\bar d_i$ &27.0885 &-56.0859 &71.2225 &2.9715 &6.61433\\[1ex] \hline
  \end{tabular}
  \caption{Parametrisations of Polyakov-loop potential in \cref{eq:polyakov_xT} and \cref{eq:polyakov_bT} .}
  \label{tab:polyakov}
\end{table}

%
\begin{figure}[b]
	\includegraphics[width=0.95\textwidth]{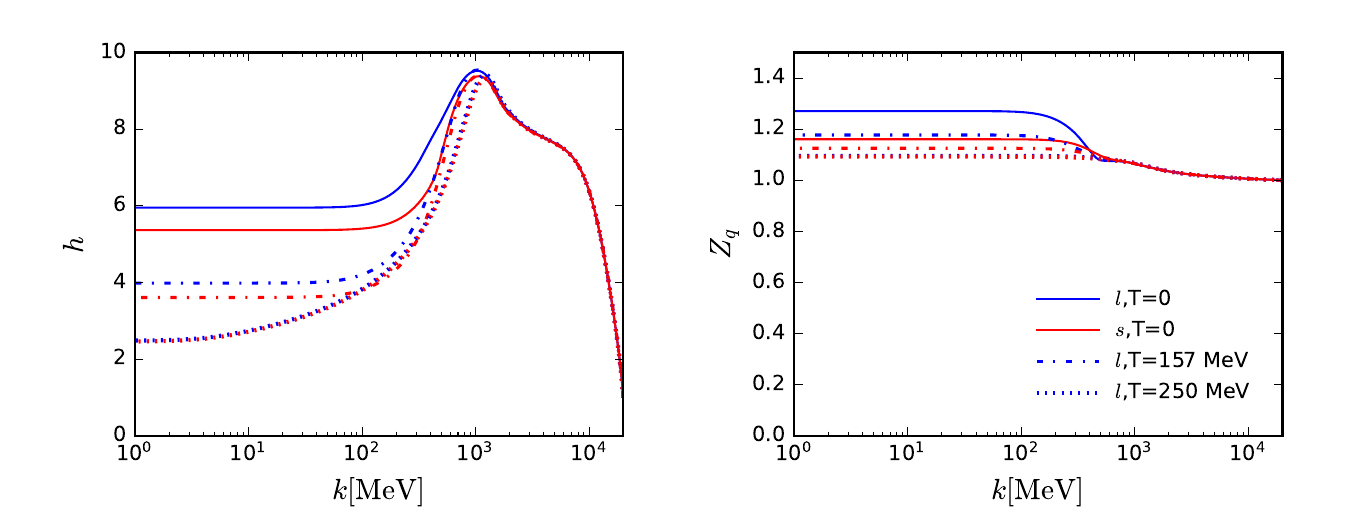}
	\caption{Light and strange quark Yukawa couplings (left panel) and quark wave functions (right panel) as functions of the RG scale $k$ for various temperatures.
	}\label{fig:hls}
\end{figure}
%

%
\begin{figure*}[t]
	\includegraphics[width=0.95\textwidth]{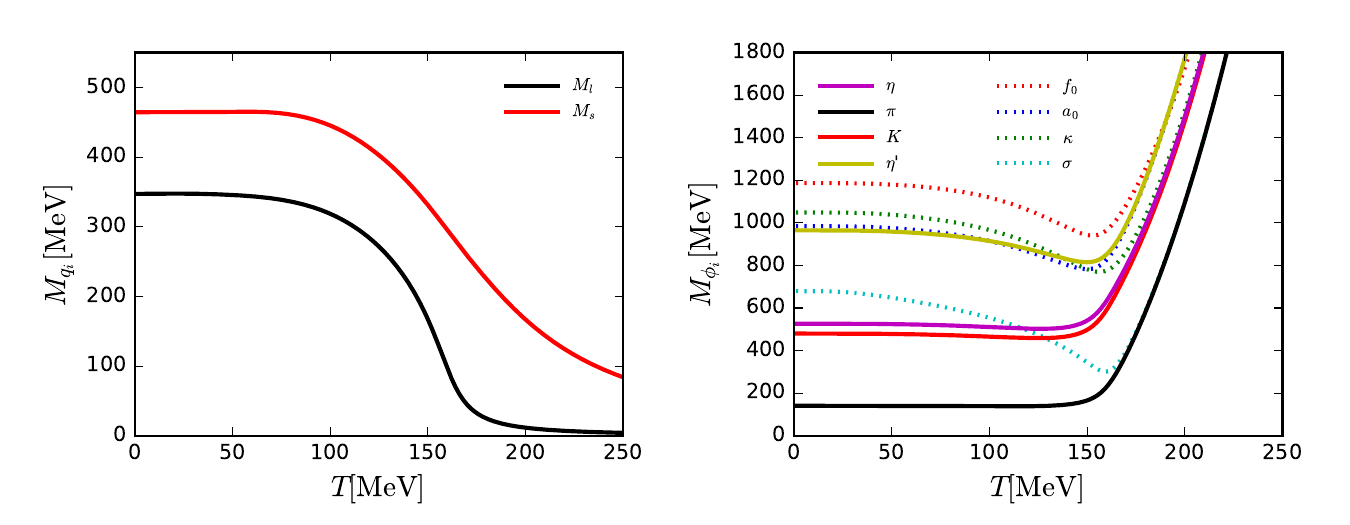}
	\caption{Constituent quark masses (left panel) and curvature meson masses (right panel) as functions of the temperature at vanishing baryon chemical potential. The solid and dashed lines in the right panel correspond to the pseudoscalar and scalar meson fields, respectively.}
	\label{fig:MFMBO}
\end{figure*}
%

\subsection{Further results} 
\label{app:Results}

In this supplement we compile further results at vanishing (\Cref{app:results_vanishing})and finite density (\Cref{app:results_finite}).

\subsubsection{Vanishing baryon chemical potential} 
\label{app:results_vanishing}

%
\begin{figure}[b]
\includegraphics[width=0.45\textwidth]{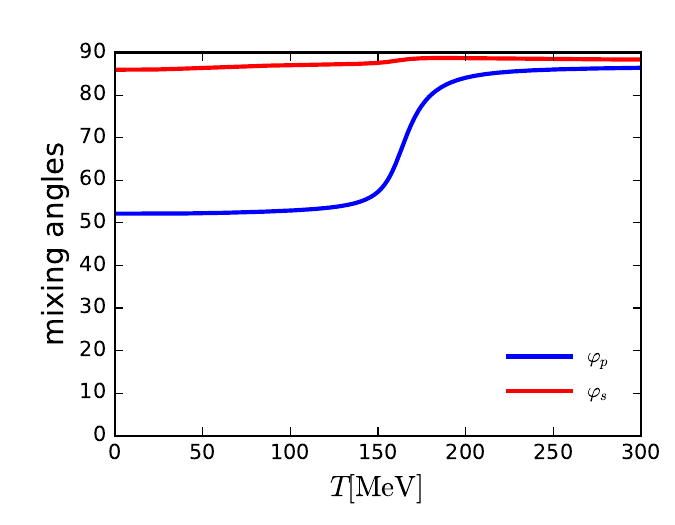}
\caption{Mixing angles between the light-strange basis and the physical basis as functions of temperature.}
\label{fig:MixingAngles}
\end{figure}
%

In the left panel of \Cref{fig:hls}, we show the light and strange Yukawa couplings as functions of the RG scale $k$. Note that our results at $k=0$ are independent of the choice of initial values for the Yukawa couplings, for more discussions see \cite{Braun:2014ata}. The light and strange Yukawa couplings split around the scale at which chiral symmetry breaks, and become degenerate again at high temperature. The quark wave functions are plotted in the right panel of \Cref{fig:hls}. In the vacuum the light and strange quark wave functions deviate in the regime of low energy, with the light wave function being slightly larger than the strange one. This result is consistent with those in \cite{Fu:2025hcm}. At high temperature, the light and strange wave functions become degenerate.

In the left panel of \Cref{fig:MFMBO}, we show the light and strange constituent quark masses, i.e. $M_{q_i}=m_{q_i}(k=0)$, as functions of the temperature at vanishing chemical potential. In the vacuum they are $M_{l}=347$ MeV and $M_{s}=464$ MeV, respectively. At high temperature, since the chiral symmetry is restored, the light quark mass tends to be zero, and the strange quark mass also decreases. The crossover temperature from the transition of quark mass is around $150$ MeV, and a precise determination based on the quark condensate is presented in the main text.

In the right panel of \Cref{fig:MFMBO}, we also show the masses of the nonets of scalar and pseudoscalar mesons, i.e. $M_{\phi_i}=m_{\phi_i}(k=0)$, as functions of the temperature. The solid and dashed lines correspond to the pseudoscalar and scalar meson fields, respectively. At zero temperature one finds $m_{\pi}=140$ MeV, $m_{K}=480$ MeV, $m_{\eta}=526$ MeV, and $m_{\eta'}=965$ MeV, which agree within our error margin with the data in the PDG \cite{Zyla:2020zbs}. 
In comparison to results in the low-energy effective theory, meson masses obtained in QCD increase faster at high temperature, indicating the rapid decoupling of mesonic degrees of freedom in QCD due to the dominating quark-gluon dynamics for large temperature in contradistinction to the lack of mesonic decoupling in low energy effective theories. 

In \Cref{fig:MFMBO}, we find that the chiral partners $\pi-\sigma$, $\eta' - a_0$, $\kappa-K$ become degenerate at high temperature. This has also been observed in the low-energy effective theories, e.g.~\cite{Schaefer:2008hk, Tiwari:2013pg, Mitter:2013fxa, Rennecke:2016tkm, Wen:2018nkn, Rai:2018ufz}. Remarkably, it is found in our calculations that the $\eta' - a_0$ is the chiral partner in QCD rather than the $\eta - a_0$. This is in agreement with the conclusion obtained from the quark-meson model with the local potential approximation (LPA) within the fRG approach \cite{Mitter:2013fxa, Rennecke:2016tkm, Wen:2018nkn} and with the mean-field approximation \cite{Schaefer:2008hk, Tiwari:2013pg, Rai:2018ufz}, while the opposite result is obtained from the quark-meson model beyond the LPA within the fRG approach \cite{Rennecke:2016tkm}. To provide a thorough analysis, we plot the mixing angles between the light-strange basis and the physical basis, defined in \Cref{eq:mix_phi}, as functions of the temperature in \Cref{fig:MixingAngles}. Obviously, the $\varphi_s$ is very close to $90^\circ$, almost being independent of the temperature, which indicates that $f_0$ is almost strange state and $\sigma$ is light state, so at high temperature, the $\pi-\sigma$ are the chiral partners. For the pseudoscalar mesons, the $\varphi_p$ is close to $54^\circ$ for $T\lesssim T_c$, which implies that $\eta$ is almost a octet state and $\eta'$ is the singlet state, see \cite{Rennecke:2016tkm}. At high temperature, the $\varphi_p\rightarrow 90^\circ$, same as the scalar case, $\eta'$ then becomes the light state and thus is the chiral partner of $a_0$.

%
\begin{figure}[t]
	\includegraphics[width=0.45\textwidth]{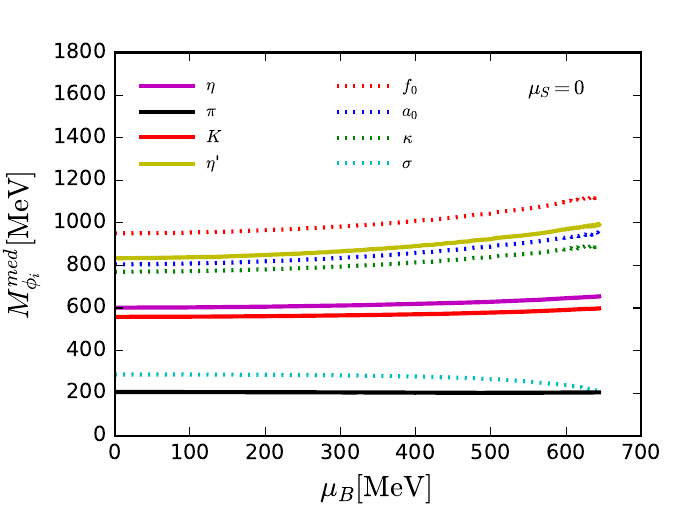}
	\caption{Meson masses as functions of the chemical potential at the pseudo-critical temperature at $\mu_S=0$.}\label{fig:MesonMedium_muS0}
\end{figure}
%

%
\begin{figure}[t]
	\includegraphics[width=0.49\textwidth]{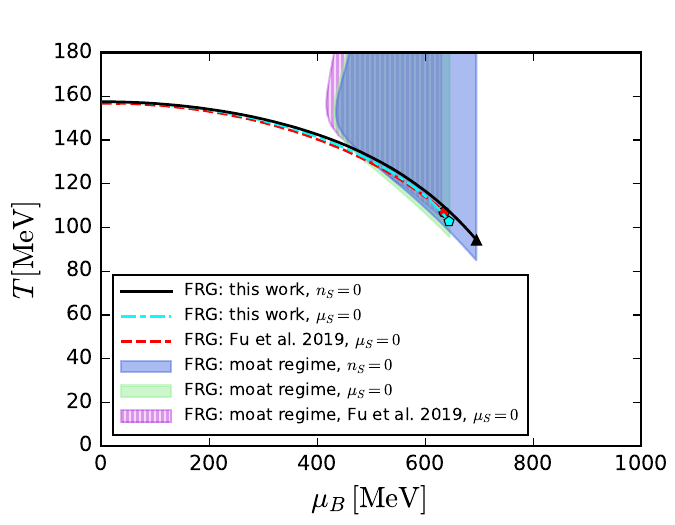}
	\includegraphics[width=0.49\textwidth]{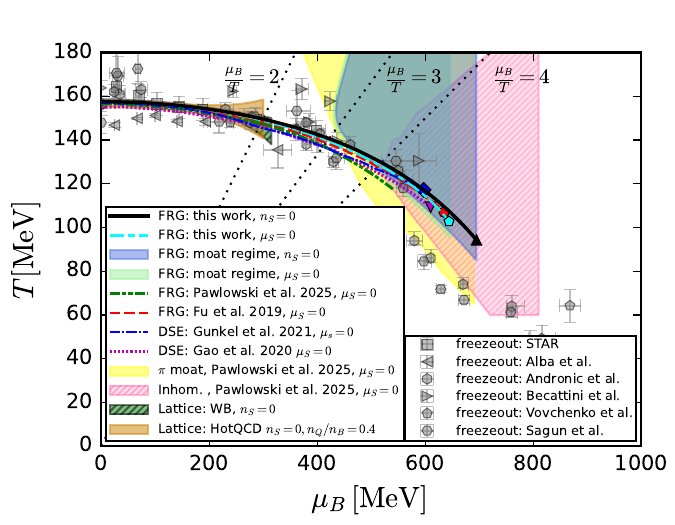}
	\caption{Left: Comparison of the QCD phase boundaries and moat regime regions between this work and \cite{Fu:2019hdw}.\\
		Right: Summary of all available data on the phase structure from fRG \cite{Fu:2019hdw, Pawlowski:2025jpg}, DSE \cite{Gao:2020fbl, Gunkel:2021oya}, and lattice QCD \cite{HotQCD:2018pds, Borsanyi:2020fev}. Freeze-out points \cite{STAR:2017sal, Alba:2014eba, Andronic:2017pug, Becattini:2016xct, Vovchenko:2015idt, Sagun:2017eye} are also plotted for comparison.}\label{fig:QCD_Phasediagram5}
\end{figure}
%
\subsubsection{Some results at finite baryon chemical potential} 
\label{app:results_finite}

For the evaluation of soft modes, in particular in the moat regime, we define the in-medium meson masses 
\begin{align}
	M_{\phi_i}^{\textrm{med}}\equiv z_{\phi_i} \lim\limits_{k \to 0} m_{\phi_i}(p=k)\,,\quad \mathrm{with} \quad z_{\phi_i} =\sqrt{\lim\limits_{k \to 0}\frac{ Z_{\phi_i} (p=k)}{Z_{\phi_i} (p=0)}}\Bigg{|}_{(T=T_{pc},\mu_B=0)}\,,
	\label{mesonmed-def}
\end{align}
which agrees with the thermal masses at $\mu_B=0$:   $M_{\phi_i}^{\textrm{med}}=M_{\phi_i}$, for further discussions see \cite{Fu:2019hdw}. 

In \Cref{fig:MesonMedium_muS0}, we show the in-medium meson masses along the chiral crossover line at $\mu_S=0$. This has to be compared with the respective behaviour for $n_S=0$, see \Cref{fig:MesonMedium}: we conclude that the behaviour of the masses is similar. Moreover, in both cases only the $\sigma$-mode drops close to the CEP, while other modes are stable or even increase slightly with the chemical potential.

We close this Supplement on further results at finite density with additional information concerning the phase structure. This complements \Cref{fig:QCD_phasediagram} in the main text: In the left panel of \Cref{fig:QCD_Phasediagram5}, we show a comparison plot of the QCD phase boundaries and moat regime regions between our results and those of \cite{Fu:2019hdw}. This allows for a further assessment of the approximations in both works as the present work builds and improves on the approximation in \cite{Fu:2019hdw}. The phase boundaries at $\mu_S=0$ agree quantitatively which each other, including the location of the CEP, which is very sensitive to any change of physics. This corroborates the results in \cite{Fu:2019hdw} and also is a non-trivial reliability check for the present work. This good agreement also extends to the shape of the moat regime region. 

Finally, in the right panel of \Cref{fig:QCD_Phasediagram5}, we provide a comprehensive Figure with all results on the phase structure with state of the art functional methods, \cite{Fu:2019hdw, Gao:2021wun, Gunkel:2021oya, Pawlowski:2025jpg}. In comparison to \Cref{fig:QCD_phasediagram}, we also include the moat regime and the inhomogeneous instabilities regions from \cite{Pawlowski:2025jpg}. \Cref{fig:QCD_Phasediagram5} summarizes the currently available results on the QCD phase structure from state-of-the-art functional and lattice QCD studies. Notably, in \cite{Pawlowski:2025jpg} the full spatial momentum dependence of the pion and $\sigma$ modes is taken into account. As a result, the moat regime found there is larger than in our work, and inhomogeneous instabilities appear in the phase diagram.


\end{document}